# Thermodynamic ($p$, $\rho$, $T$) characterization of a reference high-calorific natural gas mixture when hydrogen is added up to 20 % (mol/mol)

Daniel Lozano-Martín[1], Fatemeh Pazoki[2], Heinrich Kipphardt[3], Peyman Khanipour[3], Dirk Tuma[3], Alfonso Horrillo[2,4], and César R. Chamorro[2,*]

[1] GETEF, Departamento de Física Aplicada, Facultad de Ciencias, Universidad de Valladolid. Paseo de Belén, 7, E-47011 Valladolid, Spain.

[2] MYER, Departamento de Ingeniería Energética y Fluidomecánica, Universidad de Valladolid. Paseo del Cauce, 59, E-47011 Valladolid, Spain.

[3] BAM Bundesanstalt für Materialforschung und -prüfung, Richard-Willstätter-Str., 11, D-12489 Berlin, Germany.

[4] CIDAUT Fundación, Plaza Vicente Aleixandre Campos, 2, E-47151 Boecillo-Valladolid, Spain.

## Abstract

The injection of hydrogen into the natural-gas grid is an alternative during the process of a gradual decarbonization of the heat and power supply. When dealing with hydrogen-enriched natural gas mixtures, the performance of the reference equations of state habitually used for natural gas should be validated by using high-precision experimental thermophysical data from multicomponent reference mixtures prepared with the lowest possible uncertainty in composition. In this work, we present experimental density data for an 11-compound high-calorific (hydrogen free) natural gas mixture and for two derived hydrogen-enriched natural gas mixtures prepared by adding (10 and 20) mol-% of hydrogen to the original standard natural gas mixture. The three mixtures were prepared gravimetrically according to ISO 6142-1 for maximum precision in their composition and thus qualify for reference materials. A single-sinker densimeter was used to determine the density of the mixtures from (250 to 350) K and up to 20 MPa. The experimental density results of this work have been compared to the densities calculated by three different reference equations of state for natural gas related mixtures: the AGA8-DC92 EoS, the GERG-2008 EoS, and an improved version of the GERG-2008 EoS. While relative deviations of the experimental density data for the hydrogen-free natural gas mixture are always within the claimed uncertainty of the three considered equations of state, larger deviations can be observed for the hydrogen-enriched natural gas mixtures from any of the three equations of state, especially for the lowest temperature and the highest pressures.

* Corresponding author e-mail: cesar.chamorro@uva.es. Tel.: +34 983185697. Fax: +34 983423363

## 1. Introduction

Hydrogen-enriched natural gas, H2NG or HENG, is a mixture of natural gas and hydrogen that can be used on existing natural gas infrastructures with little or even no modification to be applied in advance. The substitution, or at least blending of natural gas with hydrogen, biomethane and synthetic methane (or e-methane), is considered a sustainable technology for a gradual decarbonization of the heat and power supply, with increasing blending fractions planned for these renewable gases with natural gas [1]. In principle, hydrogen can be mixed with natural gas in any ratio, but H2NG mixtures with up to 20 vol-% of $H_2$ represent the most realistic near-term option due to technical and economic reasons [2–5].

Several issues will need to be addressed before large-scale hydrogen injection into the natural gas grid is achieved [6]. From the point of view of end-use appliances, changes in the combustion properties of the mixture, reflected in its heating value and Wobbe index, can modify the performance of the equipment [7–10]. Another important issue is hydrogen embrittlement, which can affect the mechanical properties of iron and steel pipes [11–13]. Last but not least, the addition of hydrogen to natural gas alters the thermodynamic properties of the mixture, which affects its transport and storage characteristics. The design of the processes involved in the production, transportation, and storage of natural gas and H2NG mixtures, with special mention being given to the custody transfer applications, rely on the accuracy of the volumetric and calorific thermophysical properties obtained from the reference thermodynamic models and equations of state [14,15].

Two of the most commonly used reference equations of state for natural-gas related mixtures are the AGA8-DC92 EoS [16], developed by the *American Gas Association (AGA)*, and the GERG-2008 EoS [17,18], from the *Groupe Européen de Recherches Gazières (GERG)*, that are both explicit in the Helmholtz energy. The accuracy of these reference equations of state directly depends on both the accuracy and range of the experimental data from where they were derived. Experimental thermophysical data for the pure components of natural gas and for all corresponding binary mixtures are fundamental for the development and any further improvement of the existing reference equations of state for natural gas [19]; while experimental data of reference-quality multicomponent mixtures, resembling the composition of an actual natural gas mixture, are relevant to test their performance at pipeline conditions. Consequently, high-precision experimental thermophysical data originating from reference-quality multicomponent H2NG mixtures, which are prepared with the lowest possible uncertainty in composition, are suitable to test the overall performance of the reference equations of state used for natural gas [20,21].

In this work, we present experimental density measurements for a standard 11-compound high-calorific ($H_2$-free) natural gas mixture and for two derived H2NG mixtures, obtained by adding 10 % and 20 % (mol/mol) hydrogen to the initial standard natural gas mixture. All mixtures investigated were prepared gravimetrically according to ISO 6142-1 [22] for reference quality and maximum precision in their composition. The density measurements were performed with a high-precision single-sinker densimeter over a temperature range from (250 to 350) K and up to a maximum pressure of 20 MPa. The experimental density results of this work are compared with the abovementioned three different reference equations of state for natural-gas related mixtures: the AGA8-DC92 EoS, the GERG-2008 EoS, and an improved version of the GERG EoS, which we label as “improved-GERG-2008 EoS” in this study. Experimental density measurements for multicomponent mixtures are scarce and necessary to test the performance of reference EoS.

## 2. Experimental

### 2.1. Mixture preparation

Three synthetic natural gas mixtures, designated as G 431 ($H_2$-free natural gas mixture, BAM cylinder no. 2030-200928), G 453 ($H_2$-enriched natural gas mixture, G 431 + 10 % $H_2$, 2036-201115), and G 454 ($H_2$-enriched natural gas mixture, G 431 + 20 % $H_2$, 2043-201124), were prepared at the Federal Institute for Materials Research and Testing (BAM Bundesanstalt für Materialforschung und -prüfung, Berlin, Germany). The first mixture, G 431, is an 11-compound mixture representative of a high-calorific natural gas composed mainly of methane (> 97 %). The following two mixtures, G 453 and G 454, are made by dilution of hydrogen into the first one until a nominal composition of (10 and 20) mol-%, respectively, is reached. All gas mixtures were prepared in aluminum cylinders of $V = 10$ dm$^3$ by the gravimetric procedure, according to the standard ISO 6142-1 [22], which yields the lowest uncertainty in the composition, using the pure components of purity and supplier listed in Table 1. These pure components were used without further purification to make several premixtures and dilutions in consecutive filling steps (for a detailed filling scheme see the Supplementary File), determining the mass of the gas constituents using an electronic comparator balance (Sartorius LA 34000P-0CE, Sartorius AG) and a high-precision mechanical balance (Voland HCE 25, Voland Corp.). For each mixture of the nominal target composition (i.e., G 431, G 453, and G 454) two calibration mixtures were prepared independently, so that no correlation between the sample mixture and the calibration mixtures exist. All gas mixtures were homogenized by rolling and heating after finishing gravimetric preparation, with the compositions in molar percentage, $x_i$, and corresponding expanded ($k = 2$) uncertainties in absolute terms, $U(x_i)$, given in Table 2.

Subsequent to homogenization and prior to shipment to the University of Valladolid (UVa), the composition of each mixture to be investigated was validated by Gas Chromatography (GC) at BAM on a multichannel process analyzer (Siemens MAXUM II, Siemens AG) using the corresponding calibration gases following the procedure (so-called "bracketing") described in the standard ISO 12963 [23]. Additional details of this validation procedure are given in a previously published paper [24]. The results of the analysis are reported in Table 3, together with the composition of the corresponding validation mixtures used for this purpose. The uncertainties in the concentration values of each component for the studied mixtures and the corresponding validation mixtures have been calculated using the law of propagation of uncertainty and the procedure specified in the Guide to the Expression of Uncertainty in Measurement (GUM) [25], from the purity of the constituents of the mixture, considering the mixture preparation procedure explained in detail in the supplementary material. The deviations between gravimetric composition and that from GC analysis had to be sufficiently low as to pass the criteria for certification established by BAM.

The molar mass $M$, normalized density $\rho_0$, higher heating value $V_C$, and Wobbe index $W_s$ for the three mixtures, estimated with the REFPROP 10 software [26,27] from the normalized composition and at reference conditions of 288.15 K and 0.101325 MPa, are also included in Table 2. It can be seen from these values that the G 431 mixture is a typical high-calorific natural gas mixture and that the addition of hydrogen results in a decrease in the normalized density, the higher heating value and the Wobbe index. The higher heating value per unit of volume of the $H_2$-enriched natural gas mixtures decreases by 7 % and 14 % with respect to the values for the original $H_2$-free natural gas mixture (G 431), when 10 % (G 453) or 20 % (G 454) of hydrogen is added. The variation in the Wobbe index is less pronounced, decreasing only by 2.5 % and 5 %, respectively.

The $T$, $p$ coordinate of the critical point of the $H_2$-free natural gas mixture (G 431) is (197.2 K, 5.4MPa), with the corresponding cricondentherm at (247.9 K, 3.9 MPa), and the cricondenbar at (225.7 K, 7.5 MPa). Similarly, for the mixture G 453 (G 431 + 10 % $H_2$), the critical point is at (176.1 K, 4.3 MPa), the cricondentherm at (247.4 K, 4.3 MPa), and the cricondenbar at (219.3 K, 9.2 MPa). Finally, for the mixture G 454 (G 431 + 20 % $H_2$), the critical point is at (160.2 K, 3.9 MPa) the cricondentherm at (246.9 K, 4.9

MPa), and the cricondenbar at (207.4 K, 11.8 MPa). Note that neopentane is not included in any of the mixture models used in this work, thus it had to be added to the concentration of *n*-pentane.

## 2.2. Equipment and measurement procedure description

The experimental part of this work is accomplished with a single-sinker magnetic suspension densimeter. It consists of a pressurized diamagnetic CuCrZr cell containing a monocrystalline silicon sinker of calibrated volume ($V_s$ = 226.4440 ± 0.0026 cm$^3$) surrounded by the sample gas. The buoyancy force is transmitted to an analytical microbalance (XPE205DR, Mettler Toledo GmbH) located above the cell at ambient pressure through a magnetic coupling device. This kind of setup is currently regarded as the most accurate to determine fluid density over a large range of temperatures and pressures, providing an absolute determination of the density without the need for calibration fluids. The principles of measurement were developed by Kleinrahm, Wagner, and Lösch [28–30], originally with two-sinker devices (improved accuracy, especially at low density, mainly due to compensation for adsorption effects, which may be relevant at very low density) and then with single-sinker devices (less complex but equally accurate at high density [31,32]).

The working equation is:

$$\rho_{fluid} = \frac{\phi_0 m_s + (m_{\mathrm{Ti}} - m_{\mathrm{Ta}}) + (W_{ZP} - W_{MP})/\alpha}{V_s(T,p)} \frac{1}{\phi_0} + \frac{\varepsilon_\rho}{\phi_0} \frac{\chi_s}{\chi_{s0}} \left( \frac{\rho_s}{\rho_0} - \frac{\rho_{fluid}}{\rho_0} \right) \rho_{fluid} \tag{1}$$

where the subscripts fluid, s, Ti, Ta, ZP, and MP stand for the fluid, sinker ("specific" in case of $\chi_s$ and $\chi_{s0}$), titanium and tantalum compensation masses, zero and measuring positions of the magnetic coupling, while the terms *m*, *V*, and $\rho$ denote the mass, volume, and density, respectively.

$\alpha$ is the so-called *calibration factor* determined by weighing two calibrated compensation masses of tantalum and titanium, alternatively placed in the upper pan of the microbalance by means of an automatically controlled changing device. Both compensation masses have nearly the same volume and their mass difference is close to the mass of the sinker. In this way, the balance is always operated near its zero position, avoiding measurement errors that originate from the non-linearity of the balance itself.

The magnetic coupling system is composed of two magnets separated by the upper cell wall: an electromagnet hanging from the lower hook of the balance, and a permanent magnet attached to the upper end of the sinker support inside the cell. The actual measuring procedure is carried out in a differential manner, in the zero position (ZP) the electromagnet only attracts the sinker support without lifting the sinker, whereas in the measuring position (MP) a higher force is exerted on the permanent magnet which also lifts the silicon sinker. The difference between the readings of the balance, *W*, in these two positions allows for cancelling the weights of the sinker support, the magnets, and the balance hook, which consequently minimizes systematic errors.

Due to the fact that the vertical positions of the ZP and MP are not exactly the same, together with other possible instabilities in the alignment of the magnets during the measuring procedure, the density determination should be corrected for the *force transmission error*. This perturbation is divided into two terms: the apparatus-specific effect and the fluid-specific effect. The apparatus-specific effect is accounted in Equation (1) by $\phi_0$, and determined by measuring the sinker in vacuum after each isotherm is finished. Neglecting this correcting term can lead to significant errors [33], therefore, it must always be considered. The fluid-specific effect is described in Equation (1) by the second term on the right side, and depends (a) on the specific magnetic susceptibility of the fluid $\chi_s$ (note that here, the subscript *s* stands for *specific* and not for *sinker*), (b) the so-called apparatus-specific constant $\varepsilon_\rho$, and (c) the reducing constants $\chi_{s0}$ = 10$^{-8}$ m$^3$·kg$^{-1}$ and $\rho_0$ = 1000 kg·m$^{-3}$. The value of $\varepsilon_\rho$ was estimated for our apparatus as a function of temperature and density in a previous work by two different methods [34]. Contrary to the apparatus-specific effect, the fluid-specific effect is less significant for diamagnetic fluids, but leads to relative errors of up to 3 % for paramagnetic fluids (since $\chi_s$ for paramagnetic fluids, which is temperature dependent, can be 100 times stronger compared to diamagnetic fluids) [35–37].

The pressure of the fluid is determined by two quartz crystal transducers: one for the low-pressure range from (0 to 3) MPa (Digiquartz 2300A-101, Paroscientific Inc.) and the other for the higher pressures in the range between (3 to 20) MPa (Digiquartz 43KR-HHT-101, Paroscientific Inc.). The estimated expanded ($k$ = 2) uncertainty is $U(p) = (7.5\cdot10^{-5}(p/\text{MPa}) + 4\cdot10^{-3})$ MPa for the low-pressure transducer, and $U(p) = (6.0\cdot10^{-5}(p/\text{MPa}) + 2\cdot10^{-3})$ MPa for the high-pressure transducer.

The temperature of the cell, thermostated by means of an oil thermal bath (Dyneo DD-1000F, Julabo GmbH) and an electrical heating cylinder around the cell connected to a temperature controller (MC-E, Julabo GmbH), is measured by two standard platinum resistance thermometers SPRT-25 (S1059PJ5X6, Minco Products Inc.) controlled by an AC resistance bridge (ASL F700, Automatic Systems Laboratory). The estimated expanded ($k$ = 2) uncertainty in temperature is $U(T)$ = 0.015 K.

A more detailed description of our equipment and the measurement procedure can be found in our previous papers [38,39].

### 2.3. Uncertainty of the measurements

The experimental overall expanded ($k$ = 2) uncertainty $U_T(\rho_{exp})$ for the density measurements is reported in Table 4 in both absolute and relative terms. It takes into account the contributions from the uncertainty of the density determination, $U(\rho_{exp})$, which has been thoroughly evaluated as a function of density and specific magnetic susceptibility in two previous works for our single-sinker densimeter [34,39]:

$$U\left(\rho_{\text{exp}}\right)/\left(\text{kg}\cdot\text{m}^{-3}\right) = 2.5\cdot10^{4}\cdot\chi_s/\left(\text{m}^3\cdot\text{kg}^{-1}\right) + 1.1\cdot10^{-4}\cdot\rho_{\text{exp}}/\left(\text{kg}\cdot\text{m}^{-3}\right) + 2.3\cdot10^{-2} \quad (2)$$

combined with the uncertainties from pressure, $u(p)$, temperature, $u(T)$, and composition, $u(x_i)$, following the law of uncertainty propagation [40]:

$$U_{\text{T}}\left(\rho_{\text{exp}}\right) = 2\left[u\left(\rho_{\text{exp}}\right)^2 + \left(\left.\frac{\partial\rho}{\partial p}\right|_{T,x} u(p)\right)^2 + \left(\left.\frac{\partial\rho}{\partial T}\right|_{p,x} u(T)\right)^2 + \sum_i\left(\left.\frac{\partial\rho}{\partial x_i}\right|_{T,p,x_j\neq x_i} u(x_i)\right)^2\right]^{0.5} \quad (3)$$

where the partial derivatives of the mixture density with respect to pressure and temperature are estimated with the REFPROP 10 software [26,27] using the improved GERG-2008 EoS [41–43]. The most significant term is due to $U(\rho_{exp})$, with values up to 0.05 kg·m$^{-3}$ (0.5 %), closely followed by $U(p)$, and then with minor contributions from $U(x_i)$ and $U(T)$, below 0.015 kg·m$^{-3}$ (0.01 %). The overall experimental expanded ($k$ = 2) uncertainty for the three mixtures ranges from (0.029 to 0.078) kg·m$^{-3}$, i.e., from (0.028 to 0.61) %.

## 3. Experimental results

The measured points of the three studied mixtures are represented in the $p,T$ plots of Figure 1, together with the saturation curves estimated from the improved GERG-2008 EoS, the ranges of interest for the gas industry at pipeline conditions, and the approved application ranges of the AGA8-DC92 EoS and GERG-2008 EoS. Measurements were carried out at five temperatures, (250, 275, 300, 325, and 350) K, in decreasing pressure steps of 1 MPa starting from 20 MPa down to 1 MPa. Tables 5, 6, and 7 list the experimental ($p$, $\rho$, $T$) data for the G 431 ($H_2$-free high calorific natural gas), G 453 (G 431 + 10 % $H_2$), and G 454 (G432 + 20 % $H_2$) mixtures, respectively, with the corresponding compositions reported in Table 2.

Tables 5, 6, and 7 also display the experimental expanded ($k$ = 2) uncertainty in density, estimated by Equation (3), in absolute terms and as a percentage, as well as the relative deviations of the experimental density from the calculated values with the AGA8-DC92 EoS, GERG-2008 EoS, and improved GERG-2008 EoS.

AGA8-DC92 EoS [16] is the reference mixture model from the *American Gas Association (AGA)*. This equation was initially formulated as a virial expansion series of the compressibility factor of fluid mixtures, and then recast as an explicit form of the Helmholtz energy to also account for the calorific properties in addition to volumetric ones [44]. Its application range is restricted to homogeneous gas and supercritical phases in the temperature ranges from (250 to 350) K and pressures up to 30 MPa.

Its – nowadays in Europe more widespread – counterpart is the GERG-2008 EoS [17,18] from the *Groupe Européen de Recherches Gazières (GERG)*, which was extended to also cover the liquid and vapor-liquid equilibria (VLE) regions, covering the ranges from (60 to 700) K and up to 70 MPa [41]. Both multiparametric models were developed to describe the thermophysical properties of natural gas mixtures, including 21 components ($C_1$ to $C_{10}$, and additionally iso-$C_4$ and iso-$C_5$, $N_2$, $CO_2$, CO, $H_2O$, $O_2$, $H_2$, Ar, He, and $H_2S$), at pipeline conditions. The accuracy of the EoS is directly determined by the availability and uncertainty of the experimental VLE, density, speed of sound, enthalpy differences, and heat capacity databases.

Two major modifications to the GERG mixture model, taking the formulation of the GERG-2008 EoS as a basic framework, were subsequently implemented based on more comprehensive and consolidated experimental data as well as on new modified fitting techniques. The application ranges of temperature, pressure, and composition were not modified. The first modification focused on meeting the demand for the accurate calculation of thermophysical properties in the subcooled liquid region between (90 to 180) K with pressures of up to 10 MPa from the liquefied natural gas (LNG) industry [42]. Thus, new binary specific departure functions for (methane + *n*-pentane) and (methane + isopentane) were developed and the departure functions for the binaries (methane + *n*-butane) and (methane + isobutane) were reparametrized. In this way, it became possible to reproduce the density measurements of LNG-type mixtures within their experimental uncertainty [45,46]; whereas the calorific properties representation was also improved with respect to the original GERG-2008 EoS.

The second modification was adopted to improve the accurate description of the thermophysical properties for hydrogen-rich multicomponent mixtures in the wider temperature, pressure, and composition ranges of processes involved in the hydrogen economy [43]. For this purpose, the first step was to switch from the original pure-fluid equations from GERG-2008 EoS to the current reference pure-fluid equations. Then, three new binary specific departure functions were developed for the binary systems ($N_2$ + $H_2$), ($CO_2$ + $H_2$), and (CO + $H_2$); whereas the original specific departure function for the binary systems ($CH_4$ + $H_2$) was reparametrized. The most significant improvement, however, was the correction of the unphysical behavior for the predicted phase envelope from the GERG-2008 EoS at temperatures lower than 120 K for the binary system ($CH_4$ + $H_2$), lower than 100 K for the ($N_2$+ $H_2$), 80 K for the (CO + $H_2$), and lower than 260 K for the ($CO_2$ + $H_2$). At these states, the original GERG-2008 EoS produces an open phase envelope, which means a liquid-liquid equilibrium, contradicting the experimental data available. Apart from fixing these artifacts, the differences with experimental VLE, homogeneous density and speed of sound data were significantly reduced in several cases or, at least, remained similar to that of GERG-2008 EoS for the remaining fluid regions. In this work, we denote the modified GERG model with the two above described adjustments as improved GERG-2008 EoS.

## 4. Discussion

### 4.1. Relative deviations of experimental data from the reference equations of state

Percentage relative deviations of experimental density data from the density calculated with the mixture models of AGA8-DC92 EoS, GERG-2008 EoS, and improved GERG-2008 EoS are depicted in Figures 2, 3, and 4, for the G 431, G 453, and G 454 mixtures, respectively.

Table 8 reports a statistical comparison of the experimental density data obtained in this work and other literature data dealing with $H_2$-enriched multicomponent mixtures [24,47] with respect to AGA8-DC92 EoS, GERG-2008 EoS, and improved GERG-2008 EoS. Here, *AARD* stands for the average absolute relative deviations, *BiasRD* the average relative deviation, , *RMSRD* the root mean square relative deviation, and *MaxRD* maximum relative deviation, as expressed in Eq. (4) to (7):

$$\mathrm{AARD} = \frac{1}{n}\sum_{i=1}^{n}\left|10^2\,\frac{\rho_{\mathrm{i,exp}}-\rho_{\mathrm{i,EoS}}}{\rho_{\mathrm{i,EoS}}}\right| \tag{4}$$

$$\mathrm{BiasRD} = \frac{1}{n}\sum_{i=1}^{n}\left(10^2\,\frac{\rho_{\mathrm{i,exp}}-\rho_{\mathrm{i,EoS}}}{\rho_{\mathrm{i,EoS}}}\right) \tag{5}$$

$$\mathrm{RMSRD} = \sqrt{\frac{1}{n}\sum_{i=1}^{n}\left(10^2\,\frac{\rho_{\mathrm{i,exp}}-\rho_{\mathrm{i,EoS}}}{\rho_{\mathrm{i,EoS}}}\right)^2} \tag{6}$$

$$\mathrm{MaxRD} = \max\left|10^2\,\frac{\rho_{\mathrm{i,exp}}-\rho_{\mathrm{i,EoS}}}{\rho_{\mathrm{i,EoS}}}\right| \tag{7}$$

Figure 2 shows that relative deviations between experimental density data for the $H_2$-free natural gas mixture (G 431) and any of the three EoS used for comparison are always smaller than the claimed uncertainties of the EoS ($U(\rho_{\mathrm{EoS}})$ = 0.1 %). The three EoS applied to our results represent the experimental density data very well, with maximum relative deviations around 0.05 % and AARD between 0.012 % when comparing to AGA8-DC92 and 0.032 % with respect to the GERG-2008 EoS.

Regarding the G 453 $H_2$-enriched natural gas mixture (G 431 + 10 % $H_2$), the relative deviations from the EoS, as shown in Figure 3, are larger than the EoS uncertainty $U(\rho_{\mathrm{EoS}})$ stated by the EoS only for the lowest isotherm of 250 K and pressures above 16 MPa when comparing to the AGA8-DC92 EoS; above 13 MPa when comparing to GERG-2008 EoS; and only above a pressure as low as 4 MPa when comparing to the improved GERG-2008 EoS. Maximum relative deviations of near 0.20 % can be seen with respect to any of the three EoS for the lowest temperature. The average absolute value of the relative deviation of the experimental density data with respect to the AGA8-DC92 and GERG-2008 EoS is around 0.030 %. A slightly worse agreement, however, is obtained with respect to the improved GERG-2008 EoS, with an AARD of 0.047 %. A similar behavior could be observed in the density of a binary mixture composed of methane and hydrogen (0.90 $CH_4$ + 0.10 $H_2$), performed with this same experimental technique and published in a previous work [48].

Figure 4 shows that the relative deviations of the experimental density data for the G 454 $H_2$-enriched natural gas mixture (G 431 + 20 % $H_2$), with respect to the AGA8-DC92 and the GERG-2008 EoS, are always smaller than the stated uncertainty of the EoS, except for one single point at the lowest temperature ($T$ = 250 K) and the highest pressure ($p$ = 20 MPa). The AARD value from these two EoS is around 0.030 %, similar to the value obtained for the natural gas mixture with 10 % of hydrogen (G 453). Again, a slightly worse agreement is obtained with the improved GERG-2008 EoS, with a maximum relative deviation of 0.19 % and an AARD of 0.052 %. Surprisingly, when comparing the experimental density data to the improved GERG-2008 EoS, not only some points at the lowest temperature ($T$ = 250 K) and highest pressures ($p$ > 10 MPa) have deviations larger than the stated uncertainty of the EoS, but also some points at $T$ = 275 K.

It can be observed that the relative deviations at the lowest pressures for the $H_2$-free natural gas mixture (G 431) do not exactly tend to zero, as would be expected for approximating an ideal gas behavior; while for the two $H_2$-enriched natural gas mixtures (G 453 and G 454), there is additionally a pronounced dispersion, that is, a wider range of deviation for different temperatures, in the low-pressure data ($p$ = 1 MPa). Sorption phenomena, which can slightly change the composition of the mixture and thus alter the density measurements, mainly in complex mixtures with numerous components as the three mixtures studied here, may be related with the deviation and dispersion of the lowest pressure data. The influence of adsorption and desorption on accurate density measurement of multicomponent gas mixtures was investigated in depth by Richter and Kleinrahm in [49]. The applied measurement procedure aims to minimize these sorption phenomena by evacuating and filling the cell several times and repeating some selected points within the challenging $p$, $T$ region over long periods of time, but in the experimental technique used in this work, the single-sinker densimeter, is not appropriate to quantify this effect. In any case, we consider these deviations well below the uncertainty of the experimental data, which at lower pressures and densities is relatively high.

In general, we could prove that the AGA8-DC92 EoS is the model that performs best for the $H_2$-free sample (AARD = 0.012 %); while GERG-2008 EoS is slightly better at predicting the densities for the two investigated $H_2$-enriched mixtures (AARD = 0.030 %); although the differences between these two equations are not significant. In contrast, the improved GERG-2008 EoS reproduces the data with larger deviations in comparison to the other two reference models for the three mixtures, with AARD values as high as 0.052 % and MaxRD about 0.20 % for the samples containing hydrogen. In multiparametric Helmholtz models, the fitting of a single property affects the description of the other properties, with an accuracy strongly dependent on the amount and quality of the experimental data used for the adjustment. Thus, it seems that the corrections made in the improved GERG-2008 EoS to cover the subcooled temperature range of LNG applications and to correct unphysical phase envelopes of binary mixtures with hydrogen for isotherms below 120 K have shifted the density estimations for some of the points at supercritical pipeline conditions to slightly worse values than those obtained by the original GERG-2008 EoS.

In a previous study, Hernández-Gómez et al. [24] determined densities for a 3 % $H_2$-natural gas mixture, using the same single-sinker with magnetic coupling employed here. The measurements were carried out in temperature and pressure ranges similar to the ones explored here, namely between (260 to 350) K and up to 20 MPa, estimating an experimental expanded ($k$ = 2) uncertainty ranging from (0.029 to 0.50) %. They found relative deviations that rise to 0.19 % as compared to predictions from AGA8-DC92 EoS, and to 0.29 % according to estimations from GERG-2008 EoS, at the lowest isotherm of 260 K and pressures higher than 15 MPa. Though these discrepancies are of the same order of magnitude as the ones obtained for the G 453 mixture (G 431 + 10 % $H_2$) at similar conditions, we assume that they could be caused not only by the relatively small addition of 3 % $H_2$, but by the significant amounts of $CO_2$ (4 mol-%) and $N_2$ (12 mol-%), as well as the higher concentrations of ethane, propane, and butane of that mixture. These differences could be partly explained by the deviations found for the binary mixtures of ($CH_4$ + $CO_2$) [50] and ($CH_4$ + $N_2$) [38].

Richter et al. [47] studied the density of a pipeline natural gas blended with hydrogen at three different concentrations of (0.05, 10, and 30) mol-%. The characterization was performed with a two-sinker magnetic suspension densimeter in a narrow temperature range of (273.15, 283.15, and 293.15) K and up to a pressure of 8 MPa. They estimated a low experimental expanded ($k$ = 2) uncertainty of only 0.02 % for densities larger than 3 $kg \cdot m^{-3}$. Note that the data for the mixture with 30 % of $H_2$ in the work of Richter et al. are reported in Table 8 just for completeness but not considered in the discussion of our work because, first, the density is only determined for one isotherm, 283.15 K, and second, the results show a negative deviation of 0.26 %, independently of the pressure. The authors of that work attributed this issue to the technical impossibility of a correct analysis of the mixture composition. It is tackled by fitting the mass density results to a second order virial expansion, from which an average experimental molar mass is determined, differing from the one given by the analyzed gas composition. Then, adjusted results are obtained in terms of shifted molar densities, which are calculated using the experimental value for the molar mass and the original data sets. Notably, in the limited pressure range of 8 MPa and intermediate isotherms explored by Richter et al. [47], there is a good agreement between our data for the G 453 mixture (G 431 + 10 % $H_2$) and the results of their work for the 10 % $H_2$ mixture. Both works show relative deviations of experimental data increasing as the temperature decreases, but only within a 0.05 % band, and a nearly constant discrepancy with temperature between AGA8-DC92 EoS and GERG-2008 EoS of about 0.04 %, which cancels out for pressures around 8 MPa. The differences between their work and ours are that we see negative deviations and a better performance of the GERG-2008 EoS model; while in their case, all deviations are positive and the AGA8-DC92 EoS reproduces their data in a better way. We assume this is again related to the difference in the composition between the studied mixtures, the 10 % $H_2$ mixture of Richter et al. has a lower $CH_4$ content, but a concentration of $CO_2$ five times higher and significant amounts of $C_2H_6$ and $C_3H_8$, apart from traces of heavier hydrocarbons ($C_{7+}$).

### 4.2. Isothermal compressibility

From the values of the experimental density, further information can be obtained from its derivatives by the application of thermodynamic potentials. Figure 5 illustrates the partial derivatives of experimental density as a function of pressure, $\left.\frac{\partial \rho_{\mathrm{exp}}}{\partial p}\right|_T$, and Table 9 lists the corresponding isothermal compressibility values, $\kappa_T = \frac{1}{\rho_{\mathrm{exp}}}\left.\frac{\partial \rho_{\mathrm{exp}}}{\partial p}\right|_T$, for the three studied mixtures in this work. They were obtained by a cubic spline interpolation of the measured density data sets; for this reason, the values at maximum and minimum pressure of each isotherm were not considered and consequently discarded. These partial derivatives show an increasingly convex shape as the temperature decreases, which flattens with increasing hydrogen content for the range of pressures and temperatures studied. The $\kappa_T$ values range from (0.0335 to 0.5361) $MPa^{-1}$ at 250 K for the $H_2$-free natural gas mixture, decreasing when both the temperature and hydrogen concentration increase.

As can be seen from the solid lines in Figure 5, there is a good agreement with the predicted $\left.\frac{\partial \rho_{\mathrm{EoS}}}{\partial p}\right|_T$ from the improved GERG-2008 EoS, and the relative deviations of experimental $\kappa_T$ from predicted values from the improved GERG-2008 EoS result in AARD of (0.18, 0.14, and 0.13) % and MaxRD of (2.0, 1.7, and 1.3) % for the $H_2$-free (G 431), 10 % $H_2$ (G 453), and 20 % $H_2$ (G 454) mixtures, respectively. In all cases, they are within the estimated experimental expanded ($k$ = 2) uncertainty of $\kappa_T$, $U_r(\kappa_T)$ = 0.7 %.

## 5. Conclusions

The performance of the reference equations of state commonly used for natural gas when applied to H2NG mixtures should be checked before its generalized use for process design and custody transfer applications. For this purpose, density measurements of three synthetic natural gas related mixtures from (250 to 350) K and pressures up to 20 MPa were performed using a high-accuracy single-sinker magnetic suspension densimeter. The first mixture is an 11-compound mixture representative of a typical high-calorific natural gas composed mainly of methane (> 97 %). The following two mixtures were elaborated by the addition of hydrogen to the first one until a nominal composition of (10 and 20) mol-%, respectively, is reached. The three mixtures were prepared gravimetrically in order to achieve the maximum possible accuracy in their composition.

The experimental density results were compared to the densities given by three different reference equations of state for natural gas related mixtures: the AGA8-DC92 EoS, the GERG-2008 EoS, and an improved version of the GERG-2008 EoS. In general, while relative deviations of the experimental density data for the hydrogen-free natural gas mixture are always within the claimed uncertainty of the three considered equations of state, larger deviations can be observed for the H2NG mixtures from any of the three equations of state, especially for the lowest temperature and the highest pressures, with maximum relative deviations of near 0.20 % with respect to any of the three EoS, above their claimed uncertainty of 0.1 %.

The conclusions obtained from this study are only valid for natural gas mixtures composed mainly of methane when hydrogen is added up to 20 %. More research is needed to evaluate the performance of these reference EoS for H2NG mixtures when the starting natural gas to which hydrogen is added has a different composition and/or the hydrogen added gives a concentration above 20 mol-%. The study of H2NG mixtures with up to 20 mol-% of hydrogen obtained by injecting hydrogen to natural gas mixtures with significant amounts of ethane, propane, butane, nitrogen, or carbon dioxide are of special interest.

## Acknowledgments

This work was funded by the European Metrology Programme on Innovation and Research (EMPIR), Funder ID: 10.13039/100014132, Grant No. 19ENG03 MefHySto; and by the Regional Government of Castilla y León (Junta de Castilla y León), the Ministry of Science and Innovation MCIN, and the European Union NextGenerationEU/PRTR, project C17.I01.P01.S21.

## Figures

**Figure 1.** *p*, *T*-phase diagram showing the experimental points measured (●) and the calculated phase envelope (solid line) using the improved GERG EoS [41–43] for: a) ($H_2$-free) natural gas mixture G 431, b) ($H_2$-enriched) natural gas mixture G 453 (G 431 + 10 % $H_2$), and c) ($H_2$-enriched) natural gas mixture G 454 (G 431 + 20 % $H_2$), respectively. The marked temperature and pressure ranges represent the range of validity of the AGA8-DC92 EoS [16] (blue dotted line) and GERG-2008 EoS [17,18] (red dashed line), and the area of interest for the gas industry (black dashed line).

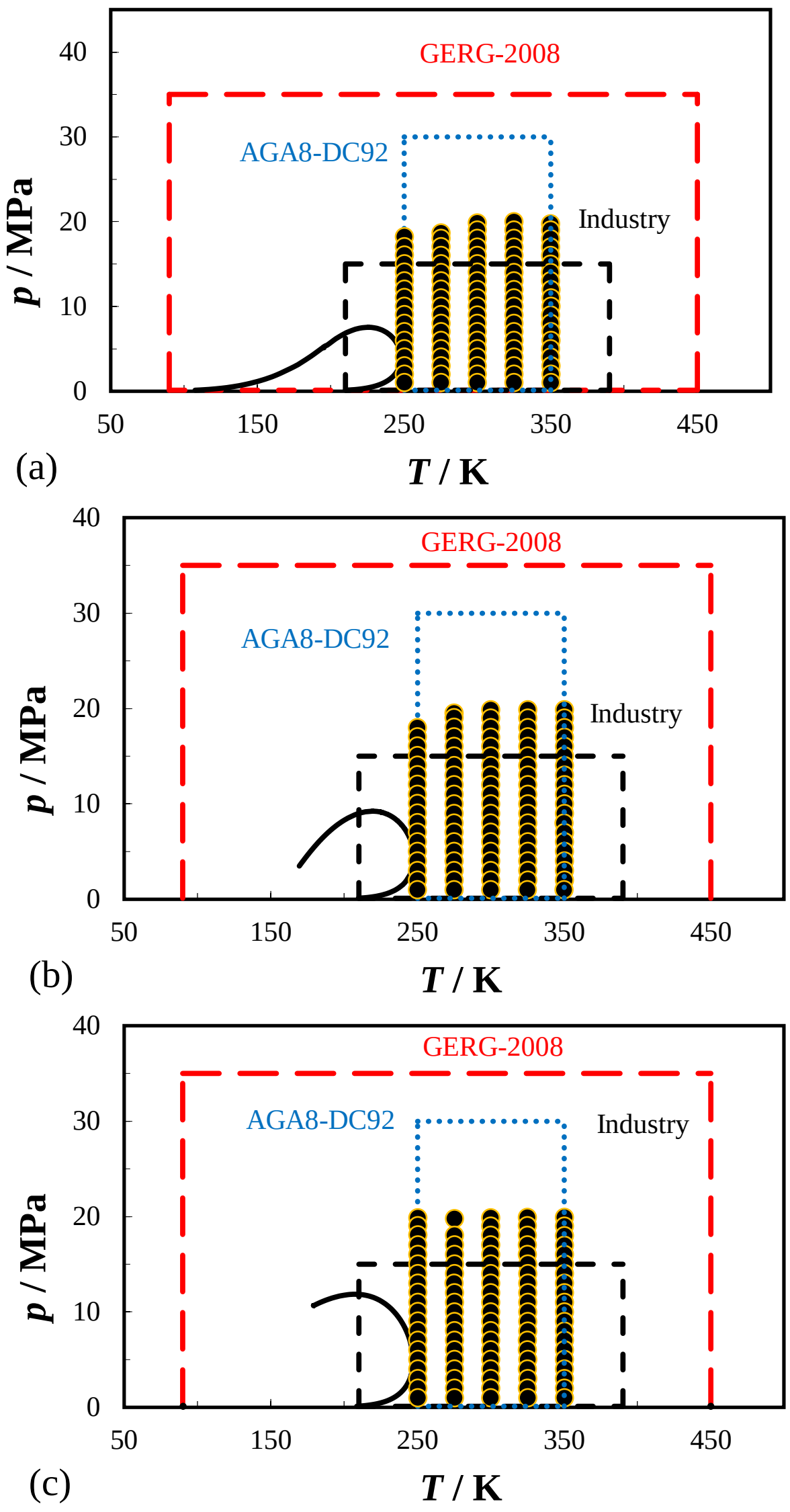

**Figure 2.** Relative deviations in density of experimental ($p$, $\rho_{exp}$, $T$) data of the ($H_2$-free) natural gas mixture G 431 from density values calculated from (a) AGA8-DC92 EoS [16], $\rho_{AGA8\text{-}DC92}$, (b) GERG-2008 EoS [17,18], $\rho_{GERG\text{-}2008}$, and (c) improved GERG-2008 EoS [41–43], $\rho_{GERG\text{-}improved}$, as a function of the pressure for different temperatures: □ 250 K, ◇ 275 K, △ 300 K, × 325 K, ○ 350 K. Dashed lines indicate the expanded ($k$ = 2) uncertainty of the corresponding EoS. Error bars on the 275-K data set indicate the expanded ($k$ = 2) uncertainty of the experimental density.

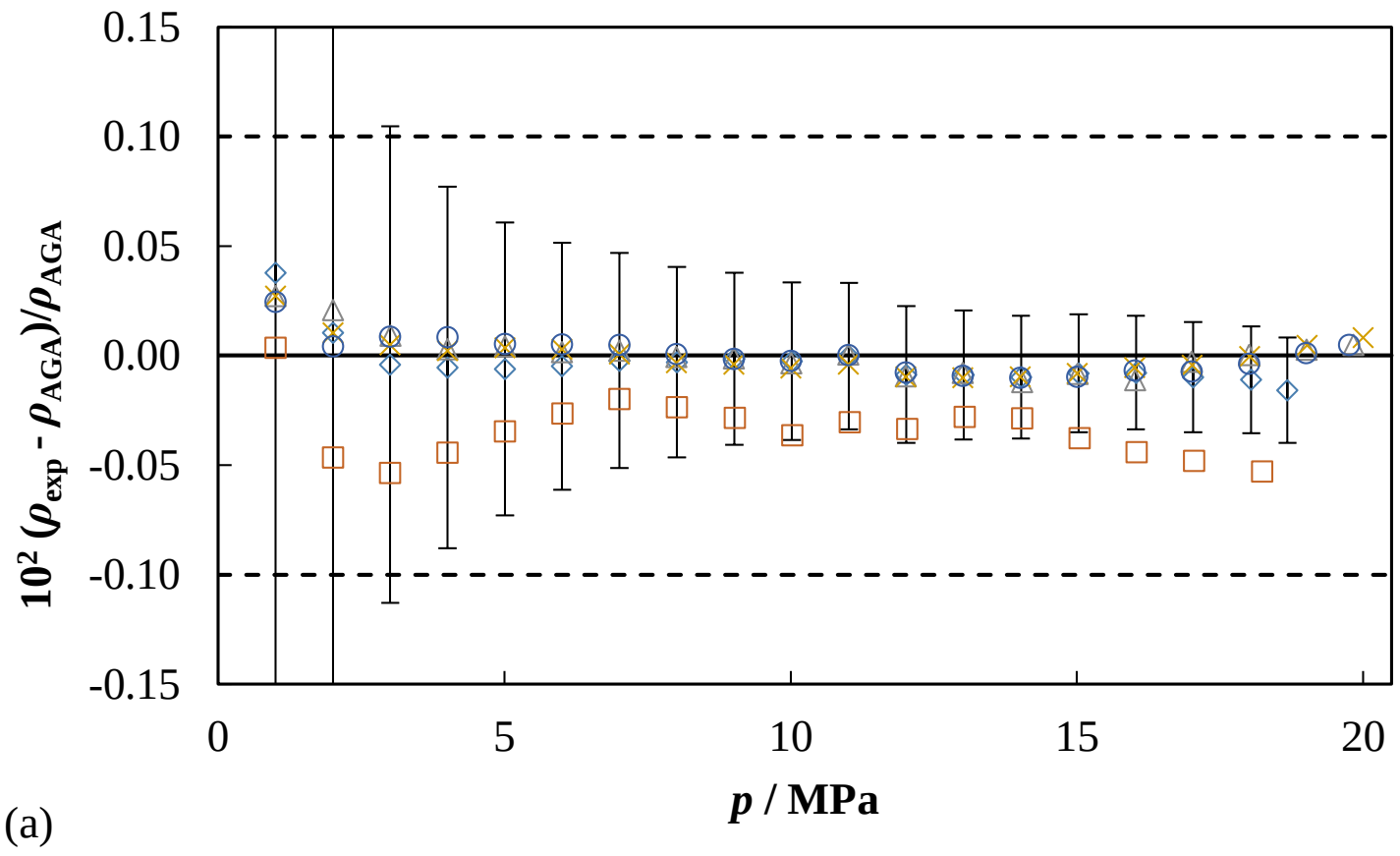

(a)

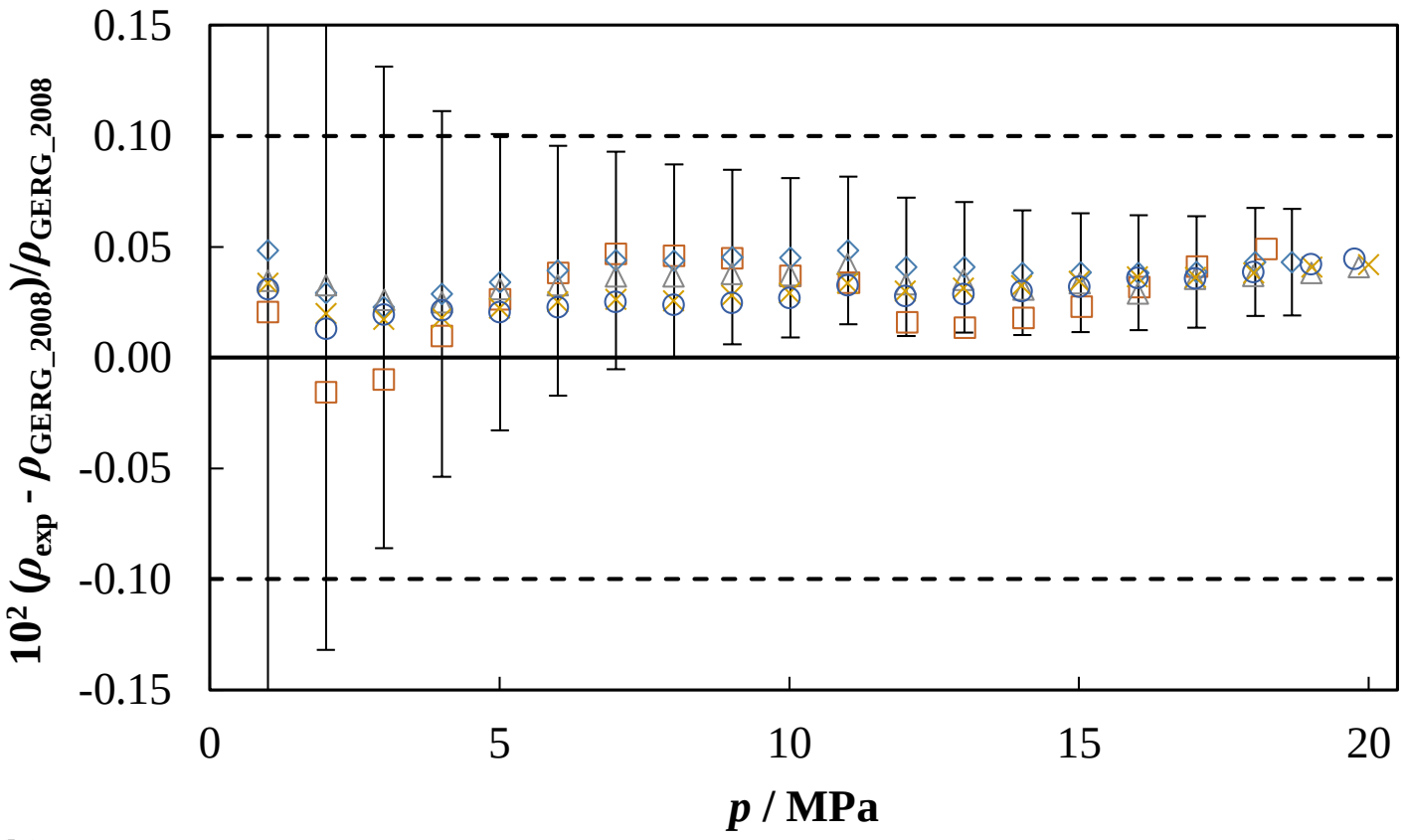

(b)

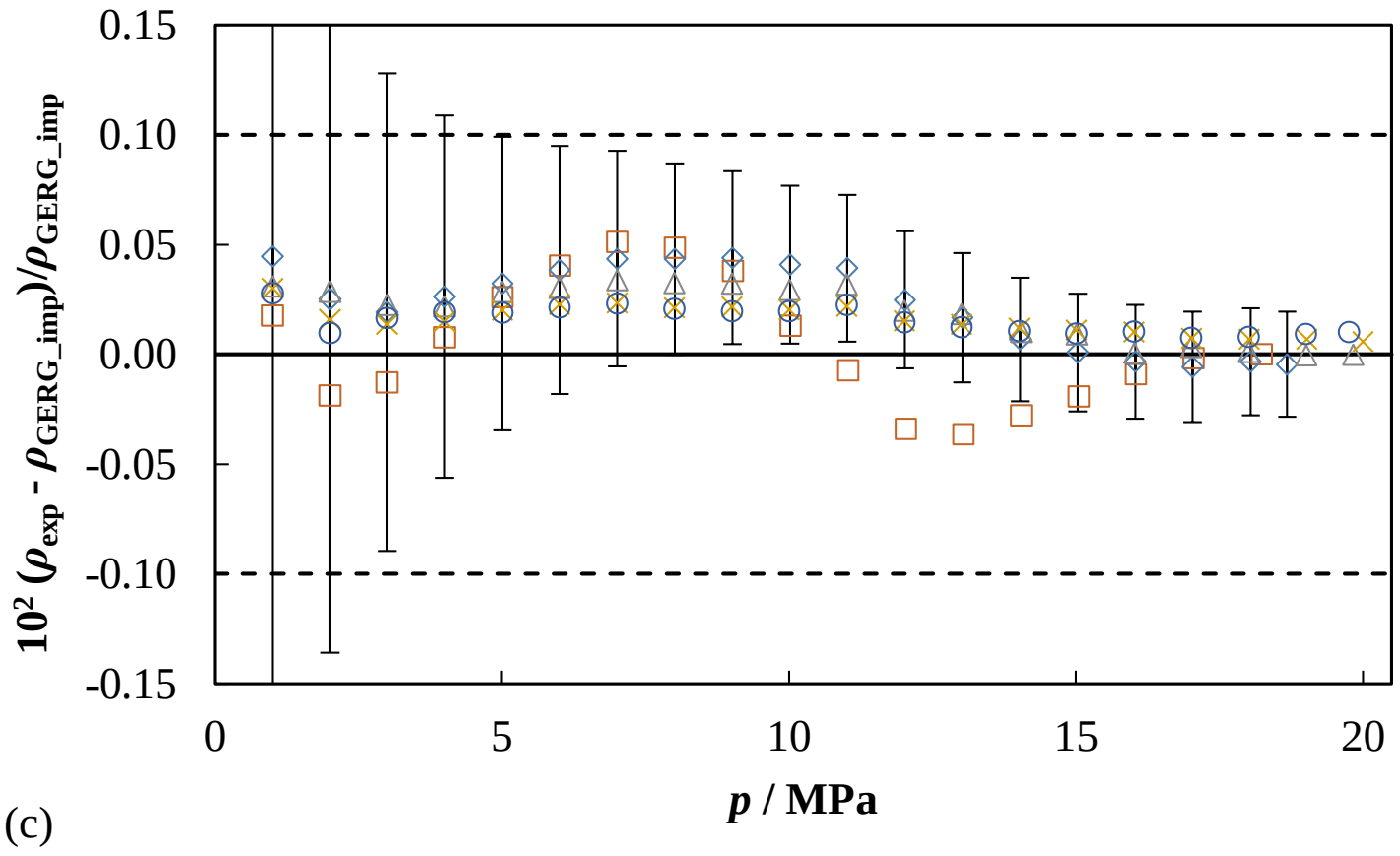

(c)

**Figure 3.** Relative deviations in density of experimental ($p$, $\rho_{exp}$, $T$) data of the ($H_2$-enriched) natural gas mixture G 453 (G 431 + 10 % $H_2$) from density values calculated from (a) AGA8-DC92 EoS [16], $\rho_{AGA8\text{-}DC92}$, (b) GERG-2008 EoS [17,18], $\rho_{GERG\text{-}2008}$, and (c) improved GERG-2008 EoS [41–43], $\rho_{GERG\text{-}improved}$, as a function of the pressure for different temperatures: □ 250 K, ◇ 275 K, △ 300 K, × 325 K, ○ 350 K. Dashed lines indicate the expanded ($k$ = 2) uncertainty of the corresponding EoS. Error bars on the 275-K data set indicate the expanded ($k$ = 2) uncertainty of the experimental density.

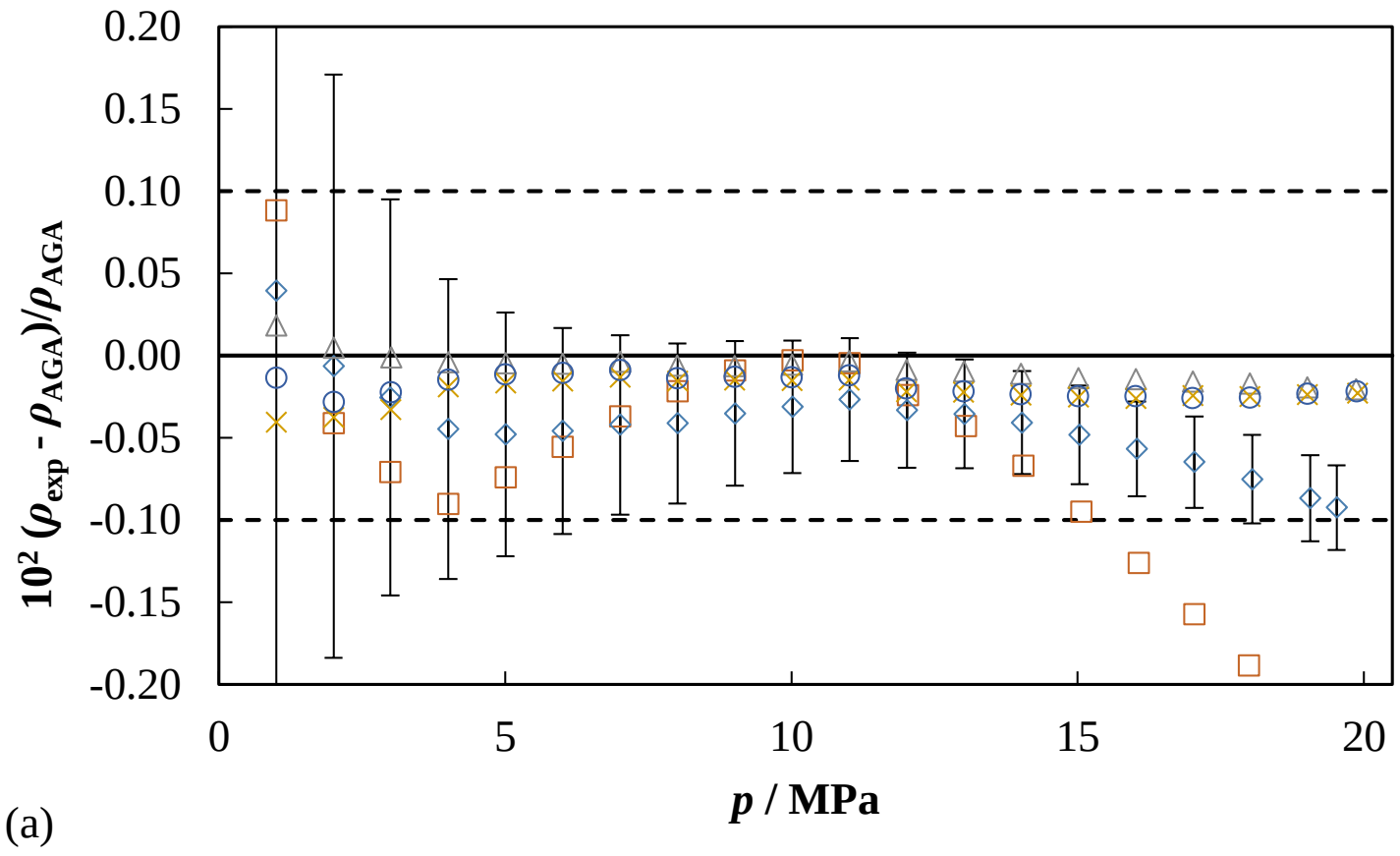

(a)

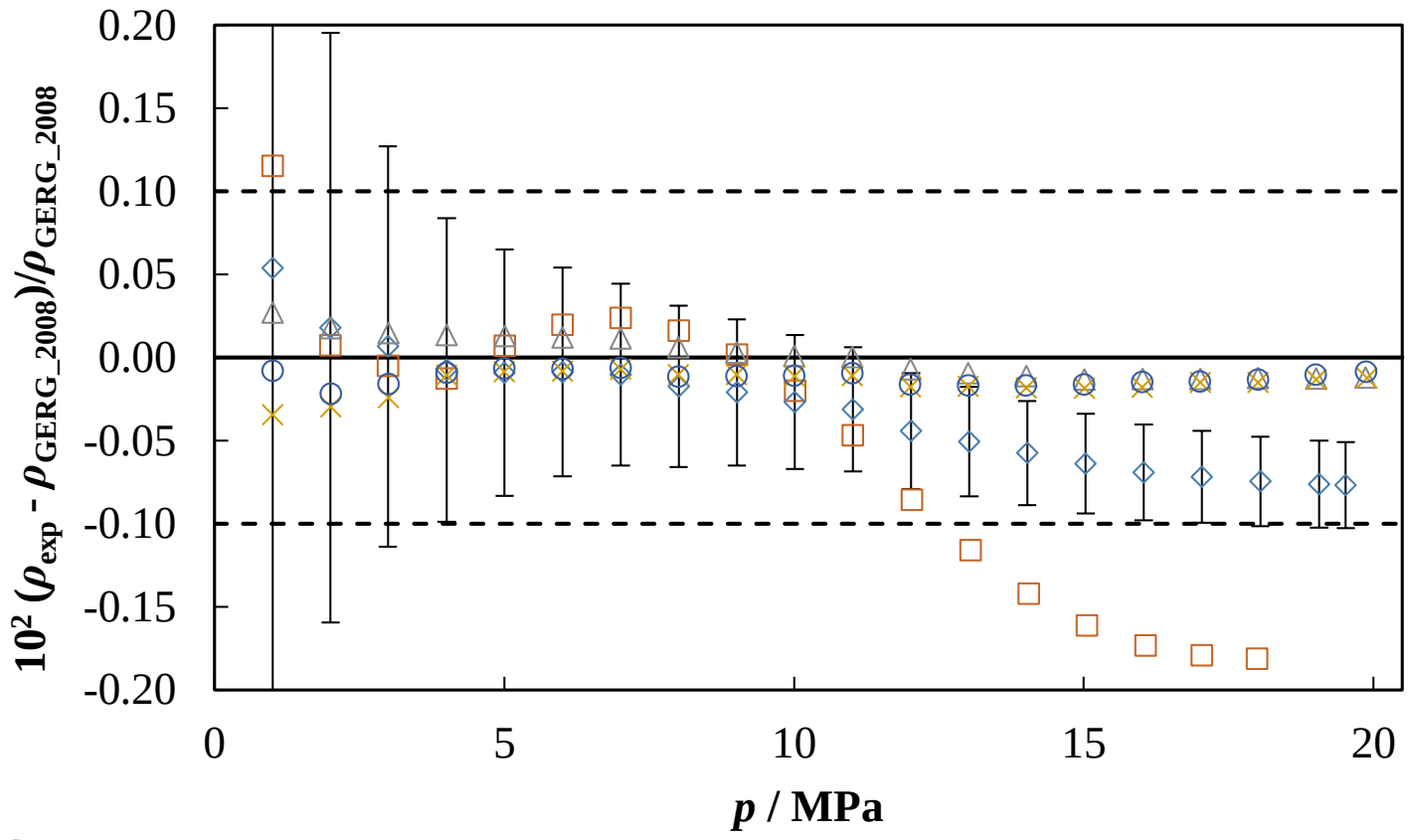

(b)

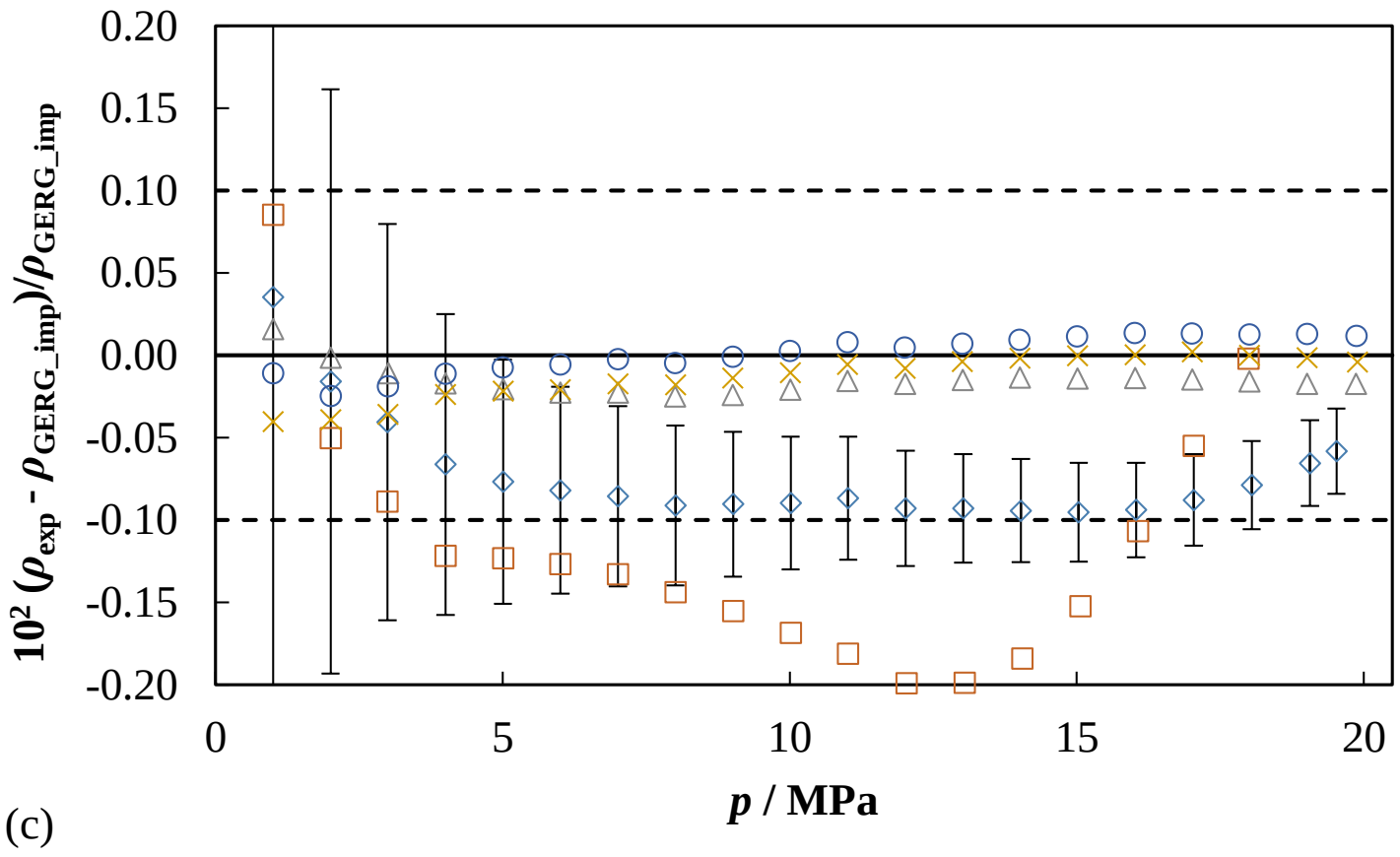

(c)

**Figure 4.** Relative deviations in density of experimental ($p$, $\rho_{exp}$, $T$) data of the ($H_2$-enriched) natural gas mixture G 454 (G 431 + 20 % $H_2$) from density values calculated from (a) AGA8-DC92 EoS [16], $\rho_{AGA8\text{-}DC92}$, (b) GERG-2008 EoS [17,18], $\rho_{GERG\text{-}2008}$, and (c) improved GERG-2008 EoS [41–43], $\rho_{GERG\text{-}improved}$, as a function of the pressure for different temperatures: □ 250 K, ◇ 275 K, △ 300 K, × 325 K, ○ 350 K. Dashed lines indicate the expanded ($k$ = 2) uncertainty of the corresponding EoS. Error bars on the 275-K data set indicate the expanded ($k$ = 2) uncertainty of the experimental density.

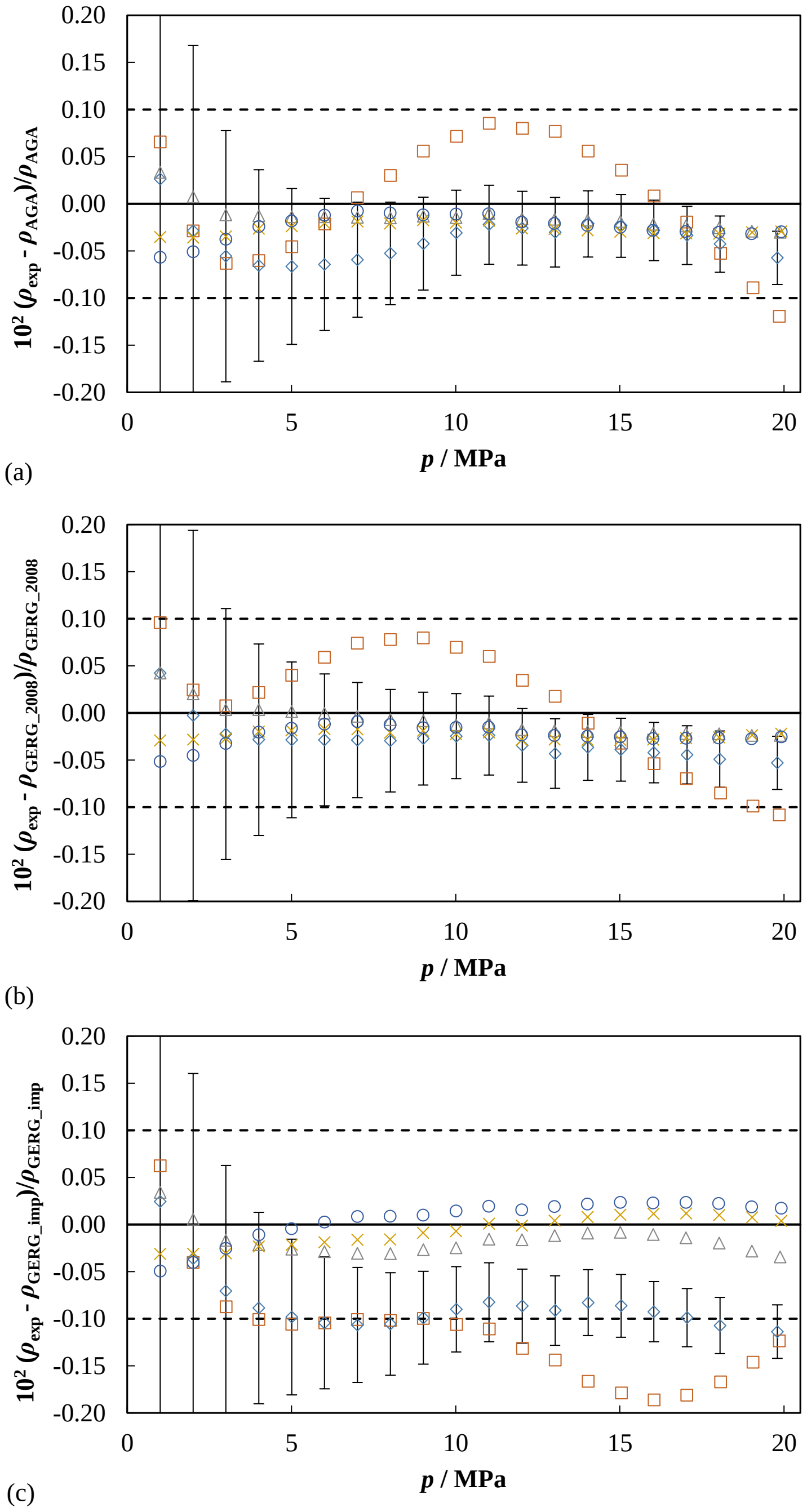

**Figure 5.** Derived $\left.\frac{\partial \rho_{\mathrm{exp}}}{\partial p}\right|_T$ values of (a) G 431, (b) G 453 (G 431 + 10 % $H_2$), and (c) G 454 (G 431 + 20 % $H_2$) mixtures as a function of the pressure for different temperatures: □ 250 K, ◇ 275 K, △ 300 K, × 325 K, ○ 350 K. Solid lines indicate the $\left.\frac{\partial \rho_{\mathrm{EoS}}}{\partial p}\right|_T$ values calculated from the improved GERG-2008 EoS [41–43].

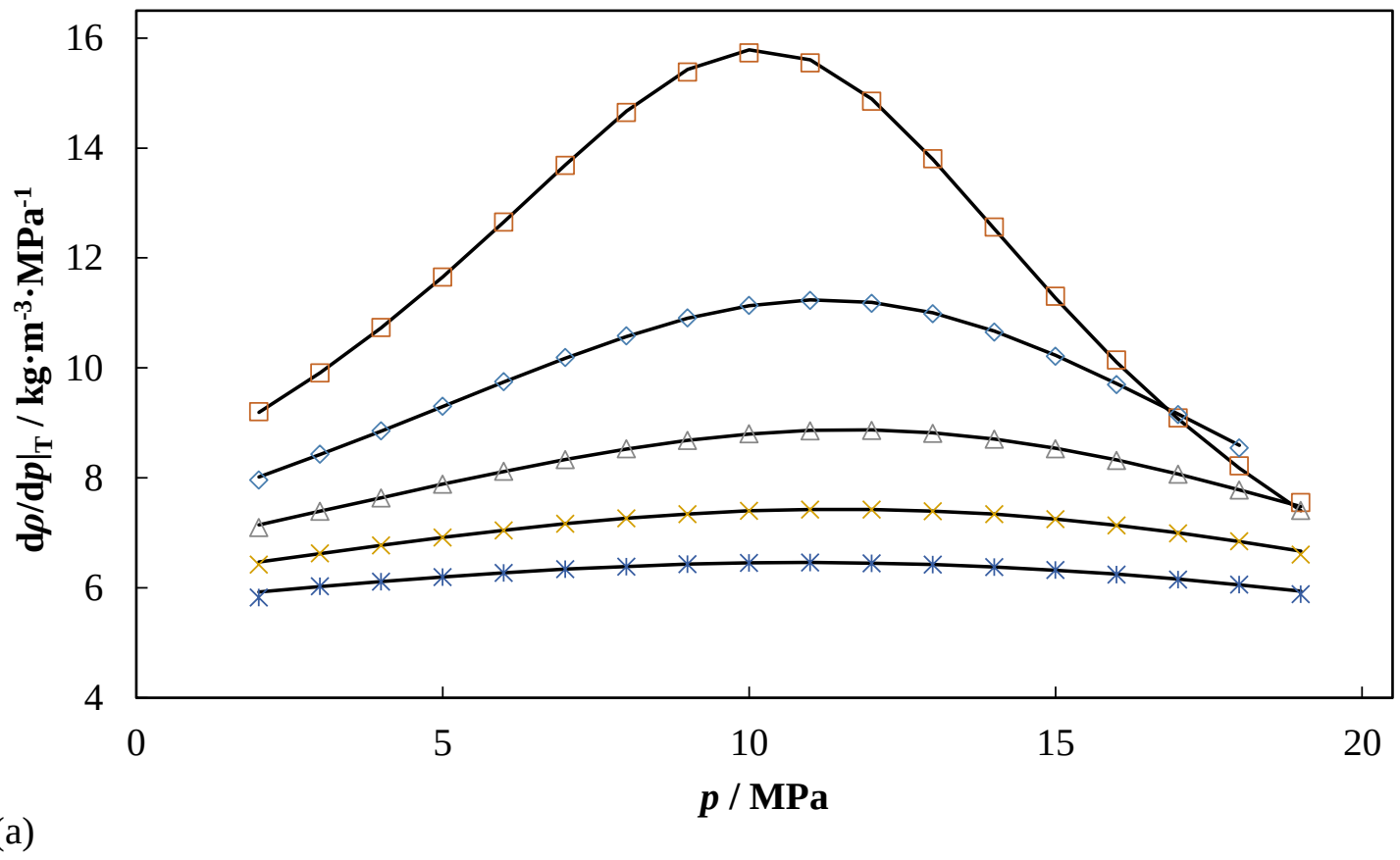

(a)

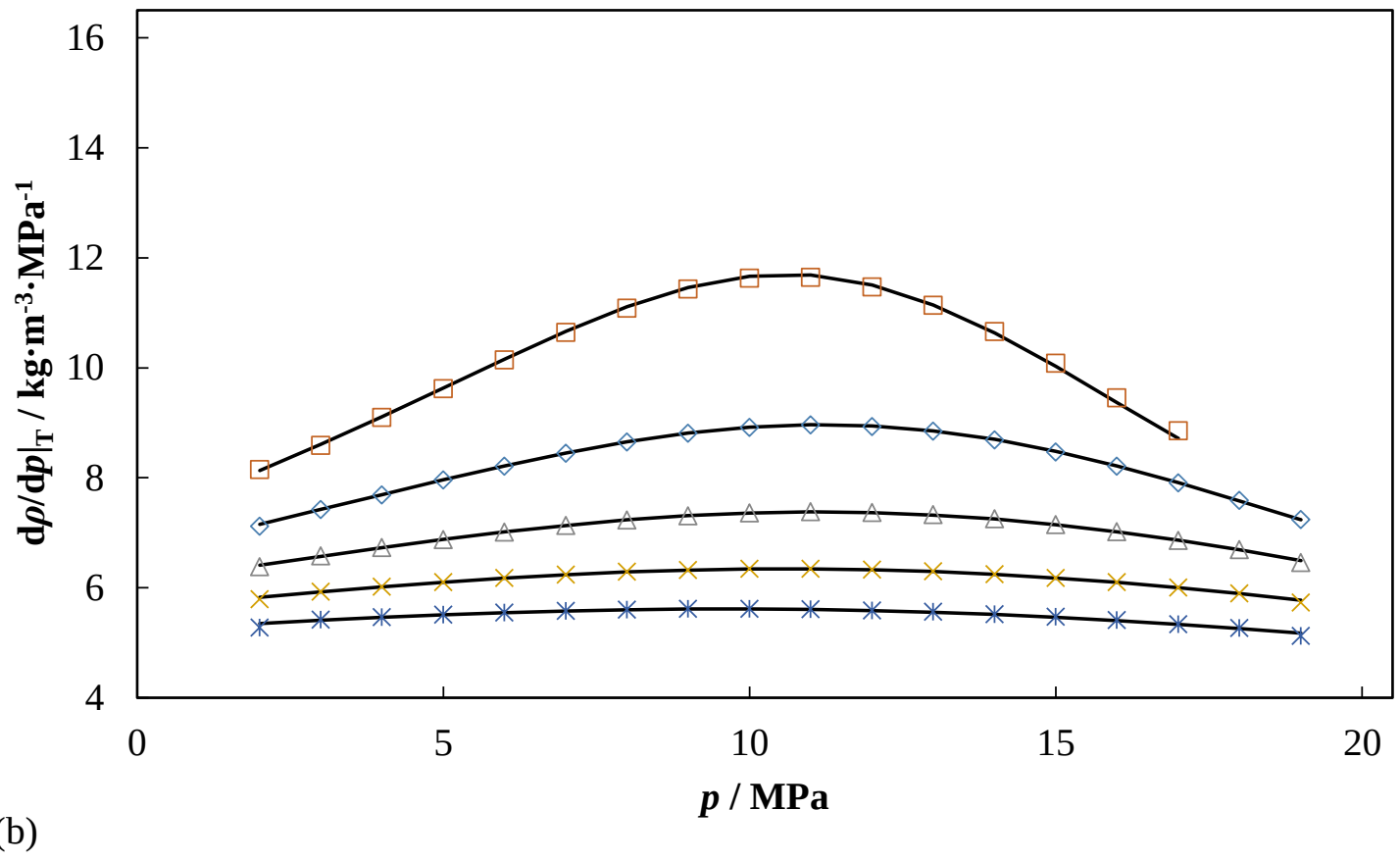

(b)

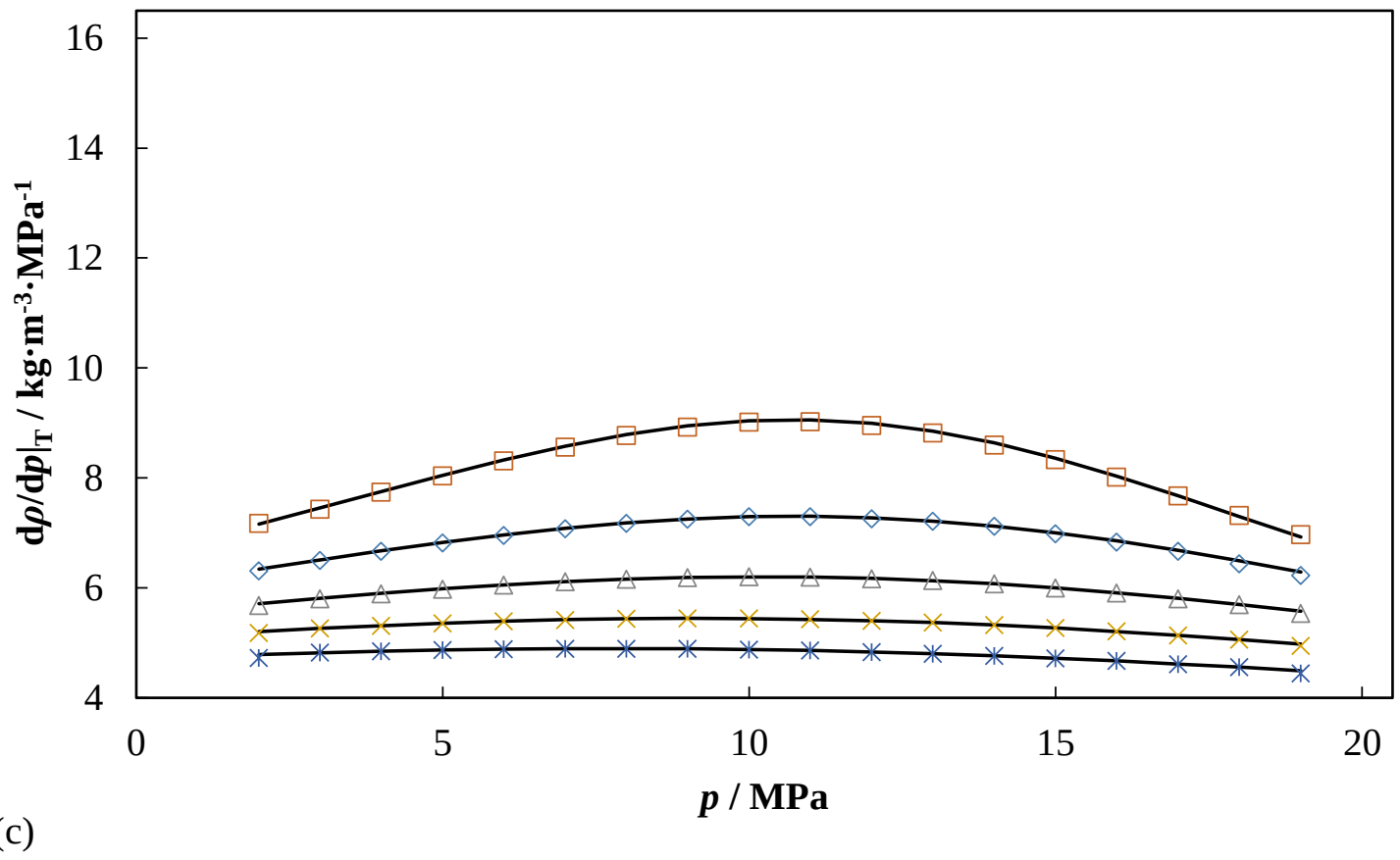

(c)

## Tables

**Table 1.** Purity, supplier, molar mass, and critical parameters for the constituting components of the reference natural gas mixtures studied in this work.

| Component (CAS no.) | Purity / mol-% | Supplier | $M$ / g·mol$^{-1}$ | Critical parameters[a] | |
|---|---|---|---|---|---|
| | | | | $T_c$ / K | $P_c$ / MPa |
| Methane (74-82-8) | ≥ 99.9995 | Linde[b] | 16.0428 | 190.56 | 4.5992 |
| Hydrogen (1333-74-0) | ≥ 99.9999 | Linde[b] | 2.01588 | 33.145 | 1.2964 |
| Nitrogen (7727-37-9) | ≥ 99.9999 | Linde[b] | 28.0135 | 126.19 | 3.3958 |
| Carbon dioxide (124-38-9) | ≥ 99.9995 | Air Liquide[c] | 44.0098 | 304.13 | 7.3773 |
| Ethane (74-84-0) | ≥ 99.999 | Matheson Tri-Gas[d] | 30.069 | 305.32 | 4.8722 |
| Propane (74-98-6) | ≥ 99.999 | Air Liquide[c] | 44.0956 | 369.89 | 4.2512 |
| *n*-Butane (106-97-8) | ≥ 99.98 | Scott[e] | 58.1222 | 425.13 | 3.796 |
| Isobutane (75-28-5) | ≥ 99.99 | Scott[f] | 58.1222 | 407.81 | 3.629 |
| *n*-Pentane (109-66-0) | ≥ 99.8 | Sigma-Aldrich[g] | 72.1488 | 469.70 | 3.3675 |
| Isopentane (78-78-4) | ≥ 99.7 | Sigma-Aldrich[g] | 72.1488 | 460.35 | 3.378 |
| Neopentane (463-82-1) | ≥ 99.0 | Linde[b] | 72.1488 | 433.74 | 3.196 |
| *n*-Hexane (110-54-3) | ≥ 99.7 | Sigma-Aldrich[g] | 86.1754 | 507.82 | 3.0441 |

[a] Critical parameters were obtained by using the default equation of state for each substance in REFPROP 10 software [26,27].

[b] Linde AG, Unterschleißheim, Germany.

[c] Air Liquide AG, Düsseldorf, Germany.

[d] Matheson Tri-Gas, Inc., Montgomeryville PA, USA.

[e] Scott UK, Newcastle-under-Lyme, UK.

[f] Scott Specialty Gases, Inc., Plumsteadville PA, USA.

[g] Sigma-Aldrich Chemie GmbH, Steinheim, Germany.

**Table 2.** Composition of the reference natural gas mixtures studied in this work. Impurity compounds are marked in *italic* type.

| Component | G 431<br>BAM no: 2030-200928 | | G 453 (G 431 + 10 % $H_2$)<br>BAM no: 2036-201115 | | G 454 (G 431 + 20 % $H_2$)<br>BAM no: 2043-201124 | |
|---|---|---|---|---|---|---|
| | $10^2\ x_i$ / mol/mol | $10^2\ U(x_i)$ / mol/mol | $10^2\ x_i$ / mol/mol | $10^2\ U(x_i)$ / mol/mol | $10^2\ x_i$ / mol/mol | $10^2\ U(x_i)$ / mol/mol |
| Methane | 97.2361 | 0.0020 | 87.5221 | 0.0023 | 77.7981 | 0.0023 |
| Hydrogen | – | – | 9.9928 | 0.0031 | 19.9945 | 0.0046 |
| Nitrogen | 1.40097 | 0.00028 | 1.25838 | 0.00024 | 1.11842 | 0.00027 |
| Carbon dioxide | 0.361460 | 0.000113 | 0.324953 | 0.000092 | 0.288529 | 0.000084 |
| Ethane | 0.398705 | 0.000033 | 0.360839 | 0.000031 | 0.320307 | 0.000031 |
| Propane | 0.201221 | 0.000020 | 0.180328 | 0.000019 | 0.160279 | 0.000019 |
| *n*-Butane | 0.100398 | 0.000052 | 0.090235 | 0.000047 | 0.079774 | 0.000042 |
| Isobutane | 0.100431 | 0.000023 | 0.090531 | 0.000021 | 0.079823 | 0.000019 |
| *n*-Pentane | 0.050072 | 0.000023 | 0.045181 | 0.000021 | 0.040107 | 0.000019 |
| Isopentane | 0.049928 | 0.000023 | 0.045043 | 0.000021 | 0.040017 | 0.000019 |
| Neopentane | 0.050781 | 0.000022 | 0.044754 | 0.000020 | 0.039945 | 0.000018 |
| *n*-Hexane | 0.049883 | 0.000018 | 0.044883 | 0.000016 | 0.040092 | 0.000015 |
| *Oxygen* | *0.000012* | *0.000012* | *0.000014* | *0.000010* | *0.000014* | *0.000011* |
| *Hydrogen* | *0.000003* | *0.000003* | – | – | – | – |
| *Carbon monoxide* | *0.0000003* | *0.0000003* | *0.000001* | *0.000001* | *0.000002* | *0.000002* |
| *Propene* | *0.0000002* | *0.0000002* | *0.0000002* | *0.0000002* | *0.0000002* | *0.0000002* |
| *Ethene* | *0.00000004* | *0.00000005* | *0.00000004* | *0.00000004* | *0.00000003* | *0.00000004* |
| *Nitric oxide* | *0.00000002* | *0.00000002* | *0.00000002* | *0.00000002* | *0.00000001* | *0.00000002* |
| Normalized composition without impurities | | | | | | |
| Methane | 97.2362 | 0.0020 | 87.5221 | 0.0023 | 77.7982 | 0.0023 |
| Hydrogen | – | – | 9.9928 | 0.0031 | 19.9946 | 0.0046 |
| Nitrogen | 1.40097 | 0.00028 | 1.25838 | 0.00024 | 1.11842 | 0.00027 |
| Carbon dioxide | 0.361460 | 0.000113 | 0.324953 | 0.000092 | 0.288529 | 0.000084 |
| Ethane | 0.398705 | 0.000033 | 0.360839 | 0.000031 | 0.320307 | 0.000031 |
| Propane | 0.201221 | 0.000020 | 0.180328 | 0.000019 | 0.160279 | 0.000019 |
| *n*-Butane | 0.100398 | 0.000052 | 0.090235 | 0.000047 | 0.079774 | 0.000042 |
| Isobutane | 0.100431 | 0.000023 | 0.090531 | 0.000021 | 0.079823 | 0.000019 |
| *n*-Pentane | 0.050072 | 0.000023 | 0.045181 | 0.000021 | 0.040107 | 0.000019 |
| Isopentane | 0.049928 | 0.000023 | 0.045043 | 0.000021 | 0.040017 | 0.000019 |
| Neopentane | 0.050782 | 0.000022 | 0.044754 | 0.000020 | 0.039945 | 0.000018 |
| *n*-Hexane | 0.049883 | 0.000018 | 0.044883 | 0.000016 | 0.040092 | 0.000015 |
| Molar mass *M* g / mol | 16.628 | | 15.167 | | 13.705 | |

| | | | |
|---|---|---|---|
| Normalized density $\rho_n$ kg / m$^3$ | 0.70468 | 0.64248 | 0.58032 |
| Higher Heating Value *HHV* MJ / m$^3$ | 37.749 | 35.174 | 32.598 |
| Wobbe index $W_s$ MJ / m$^3$ | 49.783 | 48.580 | 47.373 |

**Table 3.** Results of the gas chromatographic (GC) analysis and relative deviations between gravimetric preparation and GC analysis for the three reference natural gas mixtures studied in this work. The results are followed by the gravimetric composition of the employed validation mixtures (bracketing analysis according to ISO 12963).

| Component | Concentration | | Relative deviation between gravimetric composition and GC analysis | Concentration | | Relative deviation between gravimetric composition and GC analysis | Concentration | | Relative deviation between gravimetric composition and GC analysis |
|---|---|---|---|---|---|---|---|---|---|
| | $10^2\ x_i$ / mol/mol | $10^2\ U(x_i)$ / mol/mol | % | $10^2\ x_i$ / mol/mol | $10^2\ U(x_i)$ / mol/mol | % | $10^2\ x_i$ / mol/mol | $10^2\ U(x_i)$ / mol/mol | % |
| | G 431, BAM no.: 2030-200928 | | | G 453, BAM no.: 2036-201115 | | | G 454, BAM no.: 2043-201124 | | |
| Methane | 97.2348 | 0.0045 | −0.001 | 87.5242 | 0.0124 | 0.002 | 77.8023 | 0.0187 | 0.005 |
| Hydrogen | – | – | – | 9.9901 | 0.0111 | −0.026 | 19.9902 | 0.0184 | −0.022 |
| Nitrogen | 1.4006 | 0.0016 | −0.023 | 1.2583 | 0.0012 | −0.003 | 1.1178 | 0.0010 | −0.055 |
| Carbon dioxide | 0.3628 | 0.0019 | 0.365 | 0.3252 | 0.0014 | 0.065 | 0.2885 | 0.0019 | −0.018 |
| Ethane | 0.3989 | 0.0030 | 0.050 | 0.3611 | 0.0027 | 0.061 | 0.3207 | 0.0029 | 0.128 |
| Propane | 0.2013 | 0.0008 | 0.051 | 0.1804 | 0.0010 | 0.034 | 0.1604 | 0.0010 | 0.081 |
| *n*-Butane | 0.1004 | 0.0002 | 0.016 | 0.0903 | 0.0001 | 0.033 | 0.0799 | 0.0001 | 0.101 |
| Isobutane | 0.1004 | 0.0001 | 0.013 | 0.0905 | 0.0001 | 0.018 | 0.0799 | 0.0001 | 0.122 |
| *n*-Pentane | 0.0501 | 0.0002 | 0.087 | 0.0452 | 0.0002 | −0.018 | 0.0401 | 0.0002 | 0.086 |
| Isopentane | 0.0499 | 0.0001 | 0.032 | 0.0451 | 0.0001 | 0.081 | 0.0400 | 0.0001 | 0.050 |
| Neopentane | 0.0508 | 0.0001 | 0.006 | 0.0447 | 0.0001 | −0.026 | 0.0400 | 0.0001 | 0.122 |
| *n*-Hexane | 0.0499 | 0.0002 | 0.022 | 0.0449 | 0.0001 | 0.080 | 0.0401 | 0.0001 | 0.074 |
| | Validation mixture BAM no: 1083-191203 (lower bracket) | | | Validation mixture BAM no.: 2035-201109 (lower bracket) | | | Validation mixture BAM no.: 2042-201122 (lower bracket) | | |
| Methane | 97.3920 | 0.0022 | | 88.1236 | 0.0023 | | 78.9676 | 0.0021 | |
| Hydrogen | _ | _ | | 9.5147 | 0.0031 | | 18.9337 | 0.0045 | |
| Nitrogen | 1.31651 | 0.00027 | | 1.19824 | 0.00030 | | 1.06501 | 0.00028 | |
| Carbon dioxide | 0.342872 | 0.000114 | | 0.308391 | 0.000101 | | 0.273484 | 0.000092 | |
| Ethane | 0.377777 | 0.000034 | | 0.342841 | 0.000033 | | 0.304656 | 0.000031 | |
| Propane | 0.191152 | 0.000022 | | 0.170067 | 0.000021 | | 0.151800 | 0.000020 | |
| *n*-Butane | 0.094863 | 0.000050 | | 0.085324 | 0.000045 | | 0.075723 | 0.000040 | |
| Isobutane | 0.094977 | 0.000022 | | 0.085613 | 0.000021 | | 0.075960 | 0.000019 | |
| *n*-Pentane | 0.047448 | 0.000023 | | 0.042813 | 0.000021 | | 0.038111 | 0.000019 | |
| Isopentane | 0.047614 | 0.000023 | | 0.042761 | 0.000021 | | 0.038033 | 0.000019 | |
| Neopentane | 0.047101 | 0.000021 | | 0.042881 | 0.000020 | | 0.037935 | 0.000018 | |
| *n*-Hexane | 0.047633 | 0.000016 | | 0.042782 | 0.000016 | | 0.038001 | 0.000015 | |
| *Oxygen* | *0.000012* | *0.000012* | | *0.000014* | *0.000011* | | *0.000015* | *0.000011* | |
| *Hydrogen* | *0.000003* | *0.000003* | | _ | _ | | _ | _ | |
| *Carbon monoxide* | *0.0000003* | *0.0000002* | | *0.000001* | *0.000001* | | *0.000002* | *0.000002* | |

| | | | | | | |
|---|---|---|---|---|---|---|
| *Propene* | *0.0000002* | *0.0000002* | *0.0000002* | *0.0000002* | *0.0000002* | *0.0000002* |
| *Ethene* | *0.00000013* | *0.00000011* | *0.00000003* | *0.00000004* | *0.00000003* | *0.00000004* |
| *Nitric oxide* | *0.00000002* | *0.00000002* | *0.00000002* | *0.00000002* | *0.00000001* | *0.00000002* |
| | Validation mixture BAM no: 6098-191209 (upper bracket) | | Validation mixture BAM no.: 2037-201117 (upper bracket) | | Validation mixture BAM no.: 2044-201130 (upper bracket) | |
| Methane | 97.0985 | 0.0022 | 86.8725 | 0.0025 | 76.7016 | 0.0021 |
| Hydrogen | _ | _ | 10.5120 | 0.0033 | 20.9777 | 0.0047 |
| Nitrogen | 1.47274 | 0.00028 | 1.32675 | 0.00033 | 1.17730 | 0.00027 |
| Carbon dioxide | 0.379488 | 0.000111 | 0.342171 | 0.000097 | 0.302677 | 0.000087 |
| Ethane | 0.418538 | 0.000037 | 0.378328 | 0.000035 | 0.336258 | 0.000031 |
| Propane | 0.210275 | 0.000022 | 0.189702 | 0.000022 | 0.168337 | 0.000019 |
| *n*-Butane | 0.105241 | 0.000054 | 0.094599 | 0.000049 | 0.084057 | 0.000044 |
| Isobutane | 0.105051 | 0.000024 | 0.094881 | 0.000022 | 0.084098 | 0.000020 |
| *n*-Pentane | 0.052474 | 0.000025 | 0.047353 | 0.000022 | 0.042013 | 0.000020 |
| Isopentane | 0.052547 | 0.000025 | 0.047262 | 0.000022 | 0.041871 | 0.000020 |
| Neopentane | 0.052589 | 0.000023 | 0.047095 | 0.000021 | 0.041943 | 0.000019 |
| *n*-Hexane | 0.052533 | 0.000017 | 0.047399 | 0.000018 | 0.042069 | 0.000016 |
| *Oxygen* | *0.000012* | *0.000012* | *0.000013* | *0.000011* | *0.000015* | *0.000011* |
| *Hydrogen* | *0.000003* | *0.000002* | _ | _ | _ | _ |
| *Carbon monoxide* | *0.0000002* | *0.0000002* | *0.000001* | *0.000001* | *0.000002* | *0.000002* |
| *Propene* | *0.0000002* | *0.0000002* | *0.0000002* | *0.0000002* | *0.0000002* | *0.0000002* |
| *Ethene* | *0.00000014* | *0.00000013* | *0.00000004* | *0.00000004* | *0.00000003* | *0.00000004* |
| *Nitric oxide* | *0.00000002* | *0.00000002* | *0.00000002* | *0.00000002* | *0.00000002* | *0.00000002* |

**Table 4.** Contributions to the overall expanded ($k$ = 2) uncertainty in density, $U_T(\rho_{exp})$, for the three reference natural gas mixtures studied in this work.

| Source | Contribution ($k$ = 2) | Units | Estimation in density ($k$ = 2) | |
|---|---|---|---|---|
| | | | kg·m$^{-3}$ | % |
| | | G 431 | | |
| Temperature, $T$ | 0.015 | K | < 0.0080 | < 0.0049 |
| Pressure, $p$ | < 0.005 | MPa | (0.021–0.068) | (0.018–0.37) |
| Composition, $x_i$ | < 0.0004 | mol·mol$^{-1}$ | < 0.0012 | < 0.0014 |
| Density, $\rho$ | (0.024–0.048) | kg·m$^{-3}$ | (0.024–0.048) | (0.022–0.41) |
| Sum | | | (0.031–0.078) | (0.028–0.54) |
| | | G 453 (G 431 + 10 % $H_2$) | | |
| Temperature, $T$ | 0.015 | K | < 0.0059 | < 0.0038 |
| Pressure, $p$ | < 0.005 | MPa | (0.019–0.051) | (0.022–0.37) |
| Composition, $x_i$ | < 0.0004 | mol·mol$^{-1}$ | < 0.014 | < 0.0084 |
| Density, $\rho$ | (0.023–0.043) | kg·m$^{-3}$ | (0.023–0.043) | (0.024–0.45) |
| Sum | | | (0.030–0.064) | (0.034–0.57) |
| | | G 454 (G 431 + 20 % $H_2$) | | |
| Temperature, $T$ | 0.015 | K | < 0.0045 | < 0.0032 |
| Pressure, $p$ | < 0.005 | MPa | (0.017–0.040) | (0.021–0.37) |
| Composition, $x_i$ | < 0.0004 | mol·mol$^{-1}$ | < 0.015 | < 0.0099 |
| Density, $\rho$ | (0.023–0.041) | kg·m$^{-3}$ | (0.023–0.041) | (0.026–0.49) |
| Sum | | | (0.029–0.055) | (0.034–0.61) |

**Table 5.** Experimental ($p$, $\rho_{exp}$, $T$) measurements for the ($H_2$-free) natural gas mixture G 431, absolute and relative expanded ($k$ = 2) uncertainty in density, $U(\rho_{exp})$, relative deviations from the density given by the AGA8-DC92 EoS [16], $\rho_{AGA8\text{-}DC92}$, the GERG-2008 EoS [17,18], $\rho_{GERG\text{-}2008}$, and the improved GERG-2008 EoS [41–43], $\rho_{GERG\text{-}improved}$.

| $T$ / K[(a)] | $p$ / MPa[(a)] | $\rho_{exp}$ / kg·m$^{-3}$[(a)] | $U(\rho_{exp})$ / kg·m$^{-3}$ | $10^2\ U(\rho_{exp})/\rho_{exp}$ | $10^2\ (\rho_{exp} - \rho_{AGA8\text{-}DC92})/\rho_{AGA8\text{-}DC92}$ | $10^2\ (\rho_{exp} - \rho_{GERG\text{-}2008})/\rho_{GERG\text{-}2008}$ | $10^2\ (\rho_{exp} - \rho_{GERG\text{-}improved})/\rho_{GERG\text{-}improved}$ |
|---|---|---|---|---|---|---|---|
| 250.000 K | | | | | | | |
| 250.133 | 18.23491 | 219.356 | 0.048 | 0.022 | −0.053 | 0.049 | < 0.001 |
| 250.133 | 17.04296 | 209.224 | 0.047 | 0.022 | −0.048 | 0.041 | −0.002 |
| 250.133 | 16.04613 | 199.707 | 0.046 | 0.023 | −0.044 | 0.032 | −0.009 |
| 250.132 | 15.04764 | 189.084 | 0.044 | 0.023 | −0.038 | 0.023 | −0.019 |
| 250.133 | 14.04021 | 177.147 | 0.043 | 0.024 | −0.029 | 0.018 | −0.028 |
| 250.132 | 13.03555 | 163.969 | 0.042 | 0.025 | −0.028 | 0.013 | −0.036 |
| 250.130 | 12.03305 | 149.638 | 0.040 | 0.027 | −0.033 | 0.016 | −0.034 |
| 250.130 | 11.03005 | 134.394 | 0.038 | 0.028 | −0.030 | 0.034 | −0.007 |
| 250.130 | 10.02337 | 118.610 | 0.036 | 0.031 | −0.036 | 0.037 | 0.013 |
| 250.132 | 9.01924 | 102.948 | 0.035 | 0.034 | −0.028 | 0.045 | 0.038 |
| 250.132 | 8.01383 | 87.817 | 0.033 | 0.037 | −0.023 | 0.046 | 0.049 |
| 250.133 | 7.01058 | 73.591 | 0.031 | 0.042 | −0.020 | 0.047 | 0.051 |
| 250.132 | 6.00807 | 60.385 | 0.030 | 0.049 | −0.026 | 0.038 | 0.040 |
| 250.132 | 5.00596 | 48.212 | 0.028 | 0.059 | −0.035 | 0.026 | 0.026 |
| 250.132 | 4.00437 | 37.010 | 0.027 | 0.073 | −0.044 | 0.010 | 0.008 |
| 250.130 | 3.00344 | 26.689 | 0.026 | 0.097 | −0.054 | −0.010 | −0.013 |
| 250.130 | 2.00266 | 17.146 | 0.025 | 0.145 | −0.046 | −0.016 | −0.019 |
| 250.129 | 1.00225 | 8.287 | 0.024 | 0.287 | 0.004 | 0.021 | 0.018 |
| 275.000 K | | | | | | | |
| 275.108 | 18.67854 | 181.509 | 0.044 | 0.024 | −0.016 | 0.043 | −0.004 |
| 275.108 | 18.04554 | 176.201 | 0.043 | 0.024 | −0.011 | 0.043 | −0.003 |
| 275.106 | 17.02865 | 167.199 | 0.042 | 0.025 | −0.010 | 0.039 | −0.006 |
| 275.107 | 16.03037 | 157.800 | 0.041 | 0.026 | −0.008 | 0.038 | −0.003 |
| 275.107 | 15.02829 | 147.828 | 0.040 | 0.027 | −0.008 | 0.038 | 0.001 |
| 275.108 | 14.02385 | 137.348 | 0.039 | 0.028 | −0.010 | 0.038 | 0.007 |
| 275.108 | 13.01897 | 126.477 | 0.037 | 0.029 | −0.009 | 0.041 | 0.017 |
| 275.109 | 12.01820 | 115.381 | 0.036 | 0.031 | −0.009 | 0.041 | 0.025 |
| 275.108 | 11.01837 | 104.183 | 0.035 | 0.033 | < 0.001 | 0.048 | 0.039 |
| 275.107 | 10.01895 | 93.003 | 0.033 | 0.036 | −0.002 | 0.045 | 0.041 |
| 275.108 | 9.01383 | 81.928 | 0.032 | 0.039 | −0.002 | 0.045 | 0.044 |
| 275.107 | 8.00999 | 71.145 | 0.031 | 0.044 | −0.003 | 0.044 | 0.043 |
| 275.107 | 7.00835 | 60.751 | 0.030 | 0.049 | −0.002 | 0.044 | 0.044 |
| 275.106 | 6.00705 | 50.773 | 0.029 | 0.056 | −0.005 | 0.039 | 0.038 |
| 275.103 | 5.00540 | 41.236 | 0.028 | 0.067 | −0.006 | 0.034 | 0.032 |
| 275.105 | 4.00342 | 32.143 | 0.027 | 0.083 | −0.005 | 0.029 | 0.026 |
| 275.104 | 3.00300 | 23.504 | 0.026 | 0.109 | −0.004 | 0.023 | 0.019 |
| 275.104 | 2.00213 | 15.283 | 0.025 | 0.161 | 0.011 | 0.029 | 0.025 |
| 275.106 | 1.00227 | 7.467 | 0.024 | 0.318 | 0.038 | 0.048 | 0.045 |
| 300.000 K | | | | | | | |
| 300.069 | 19.83028 | 160.549 | 0.041 | 0.026 | 0.005 | 0.041 | < 0.001 |
| 300.066 | 19.00896 | 154.521 | 0.040 | 0.026 | 0.003 | 0.038 | < 0.001 |
| 300.061 | 18.00697 | 146.887 | 0.040 | 0.027 | < 0.001 | 0.037 | 0.001 |
| 300.060 | 17.00683 | 138.967 | 0.039 | 0.028 | −0.003 | 0.035 | 0.003 |
| 300.049 | 16.01803 | 130.872 | 0.038 | 0.029 | −0.011 | 0.029 | 0.001 |
| 300.057 | 15.01473 | 122.419 | 0.037 | 0.030 | −0.008 | 0.034 | 0.009 |
| 300.052 | 14.04309 | 114.049 | 0.036 | 0.031 | −0.012 | 0.031 | 0.010 |
| 300.059 | 13.00805 | 104.984 | 0.035 | 0.033 | −0.008 | 0.035 | 0.018 |
| 300.057 | 12.00698 | 96.131 | 0.034 | 0.035 | −0.009 | 0.033 | 0.020 |
| 300.066 | 11.00670 | 87.265 | 0.033 | 0.038 | < 0.001 | 0.042 | 0.032 |
| 300.064 | 10.00708 | 78.432 | 0.032 | 0.041 | −0.004 | 0.037 | 0.029 |
| 300.072 | 9.00658 | 69.682 | 0.031 | 0.044 | −0.001 | 0.038 | 0.032 |
| 300.074 | 8.00637 | 61.071 | 0.030 | 0.049 | < 0.001 | 0.037 | 0.033 |
| 300.073 | 7.00541 | 52.630 | 0.029 | 0.055 | 0.002 | 0.037 | 0.034 |
| 300.072 | 6.00469 | 44.394 | 0.028 | 0.063 | 0.002 | 0.033 | 0.030 |
| 300.072 | 5.00421 | 36.387 | 0.027 | 0.074 | 0.004 | 0.031 | 0.029 |
| 300.067 | 4.00390 | 28.621 | 0.026 | 0.091 | 0.003 | 0.025 | 0.022 |
| 300.068 | 3.00496 | 21.112 | 0.025 | 0.120 | 0.009 | 0.026 | 0.022 |
| 300.071 | 2.00279 | 13.828 | 0.024 | 0.177 | 0.021 | 0.033 | 0.029 |
| 300.069 | 1.00325 | 6.807 | 0.024 | 0.347 | 0.027 | 0.035 | 0.031 |
| 325.000 K | | | | | | | |
| 325.063 | 20.00253 | 140.124 | 0.039 | 0.028 | 0.008 | 0.042 | 0.006 |
| 325.065 | 19.02061 | 133.677 | 0.038 | 0.029 | 0.005 | 0.041 | 0.007 |
| 325.063 | 18.01414 | 126.885 | 0.037 | 0.029 | < 0.001 | 0.038 | 0.007 |
| 325.063 | 17.01119 | 119.948 | 0.037 | 0.030 | −0.004 | 0.036 | 0.007 |
| 325.064 | 16.00759 | 112.859 | 0.036 | 0.032 | −0.005 | 0.036 | 0.010 |
| 325.065 | 15.00686 | 105.660 | 0.035 | 0.033 | −0.008 | 0.034 | 0.011 |
| 325.065 | 14.00756 | 98.370 | 0.034 | 0.035 | −0.010 | 0.033 | 0.012 |
| 325.063 | 13.00645 | 90.995 | 0.033 | 0.037 | −0.010 | 0.031 | 0.014 |
| 325.064 | 12.00582 | 83.578 | 0.032 | 0.039 | −0.010 | 0.030 | 0.015 |
| 325.064 | 11.00642 | 76.158 | 0.032 | 0.041 | −0.004 | 0.034 | 0.022 |
| 325.063 | 10.00509 | 68.733 | 0.031 | 0.045 | −0.006 | 0.029 | 0.020 |
| 325.064 | 9.00636 | 61.370 | 0.030 | 0.049 | −0.004 | 0.028 | 0.022 |
| 325.065 | 8.00457 | 54.051 | 0.029 | 0.054 | −0.003 | 0.026 | 0.021 |
| 325.063 | 7.00428 | 46.834 | 0.028 | 0.060 | 0.001 | 0.026 | 0.023 |
| 325.064 | 6.00536 | 39.734 | 0.027 | 0.069 | 0.003 | 0.025 | 0.023 |
| 325.063 | 5.00332 | 32.737 | 0.027 | 0.081 | 0.004 | 0.022 | 0.020 |
| 325.064 | 4.00301 | 25.888 | 0.026 | 0.100 | 0.002 | 0.018 | 0.015 |
| 325.066 | 3.00225 | 19.184 | 0.025 | 0.131 | 0.005 | 0.017 | 0.014 |
| 325.067 | 2.00329 | 12.644 | 0.024 | 0.192 | 0.011 | 0.020 | 0.016 |
| 325.069 | 1.00007 | 6.234 | 0.024 | 0.378 | 0.027 | 0.034 | 0.030 |
| 350.000 K | | | | | | | |
| 350.055 | 19.75548 | 122.747 | 0.037 | 0.030 | 0.005 | 0.045 | 0.010 |
| 350.060 | 19.00598 | 118.326 | 0.036 | 0.031 | 0.001 | 0.042 | 0.009 |
| 350.062 | 18.00664 | 112.330 | 0.036 | 0.032 | −0.004 | 0.039 | 0.008 |
| 350.065 | 17.00571 | 106.218 | 0.035 | 0.033 | −0.007 | 0.036 | 0.008 |
| 350.070 | 16.00649 | 100.022 | 0.034 | 0.034 | −0.007 | 0.036 | 0.010 |
| 350.069 | 15.00496 | 93.728 | 0.034 | 0.036 | −0.010 | 0.032 | 0.009 |
| 350.069 | 14.00607 | 87.384 | 0.033 | 0.038 | −0.010 | 0.030 | 0.011 |
| 350.071 | 13.00542 | 80.977 | 0.032 | 0.040 | −0.009 | 0.029 | 0.013 |
| 350.074 | 12.00550 | 74.541 | 0.031 | 0.042 | −0.008 | 0.028 | 0.015 |
| 350.074 | 11.00471 | 68.085 | 0.031 | 0.045 | < 0.001 | 0.033 | 0.023 |
| 350.074 | 10.00535 | 61.630 | 0.030 | 0.049 | −0.002 | 0.027 | 0.020 |
| 350.073 | 9.00492 | 55.186 | 0.029 | 0.053 | −0.001 | 0.025 | 0.020 |
| 350.071 | 8.00374 | 48.769 | 0.028 | 0.058 | 0.001 | 0.024 | 0.021 |
| 350.074 | 7.00271 | 42.398 | 0.028 | 0.065 | 0.005 | 0.025 | 0.023 |
| 350.074 | 6.00345 | 36.095 | 0.027 | 0.075 | 0.005 | 0.023 | 0.021 |
| 350.072 | 5.00322 | 29.857 | 0.026 | 0.088 | 0.005 | 0.021 | 0.019 |
| 350.073 | 4.00241 | 23.695 | 0.026 | 0.108 | 0.008 | 0.021 | 0.019 |
| 350.075 | 3.00359 | 17.634 | 0.025 | 0.141 | 0.009 | 0.020 | 0.017 |
| 350.077 | 2.00208 | 11.651 | 0.024 | 0.208 | 0.004 | 0.013 | 0.010 |
| 350.080 | 1.00190 | 5.779 | 0.024 | 0.407 | 0.025 | 0.031 | 0.028 |

[a] Expanded uncertainties ($k$ = 2): $U(p > 3)/\mathrm{MPa} = 75 \cdot 10^{-6} \cdot \frac{p}{\mathrm{MPa}} + 3.5 \cdot 10^{-3}$; $U(p < 3)/\mathrm{MPa} = 60 \cdot 10^{-6} \cdot \frac{p}{\mathrm{MPa}} + 1.7 \cdot 10^{-3}$; $U(T)$ = 15 mK; $\frac{U(\rho)}{\mathrm{kg \cdot m^{-3}}} = 2.5 \cdot 10^{4} \frac{\chi s}{\mathrm{m^{3} kg^{-1}}} + 1.1 \cdot 10^{-4} \cdot \frac{\rho}{\mathrm{kg \cdot m^{-3}}} + 2.3 \cdot 10^{-2}$.

**Table 6.** Experimental ($p$, $\rho_{exp}$, $T$) measurements for the ($H_2$-enriched) natural gas mixture G 453 (G 431 + 10 % $H_2$), absolute and relative expanded ($k$ = 2) uncertainty in density, $U(\rho_{exp})$, relative deviations from the density given by the AGA8-DC92 EoS [16], $\rho_{AGA8\text{-}DC92}$, the GERG-2008 EoS [17,18], $\rho_{GERG\text{-}2008}$, and the improved GERG-2008 EoS [41–43], $\rho_{GERG\text{-}improved}$.

| $T$ / K[(a)] | $p$ / MPa[(a)] | $\rho_{exp}$ / kg·m$^{-3}$[(a)] | $U(\rho_{exp})$ / kg·m$^{-3}$ | $10^2$ $U(\rho_{exp})/\rho_{exp}$ | $10^2$ $(\rho_{exp} - \rho_{AGA8\text{-}DC92})/\rho_{AGA8\text{-}DC92}$ | $10^2$ $(\rho_{exp} - \rho_{GERG\text{-}2008})/\rho_{GERG\text{-}2008}$ | $10^2$ $(\rho_{exp} - \rho_{GERG\text{-}improved})/\rho_{GERG\text{-}improved}$ |
|---|---|---|---|---|---|---|---|
| | | | | 250.000 K | | | |
| 250.038 | 17.99536 | 178.985 | 0.043 | 0.024 | −0.189 | −0.181 | −0.002 |
| 250.037 | 17.03752 | 170.862 | 0.042 | 0.025 | −0.157 | −0.179 | −0.055 |
| 250.036 | 16.07095 | 162.081 | 0.041 | 0.026 | −0.126 | −0.173 | −0.107 |
| 250.036 | 15.06005 | 152.263 | 0.040 | 0.026 | −0.095 | −0.161 | −0.152 |
| 250.035 | 14.05220 | 141.856 | 0.039 | 0.028 | −0.067 | −0.142 | −0.184 |
| 250.037 | 13.04991 | 130.960 | 0.038 | 0.029 | −0.043 | −0.116 | −0.199 |
| 250.036 | 12.03622 | 119.509 | 0.036 | 0.031 | −0.024 | −0.086 | −0.199 |
| 250.037 | 11.01667 | 107.723 | 0.035 | 0.033 | −0.005 | −0.047 | −0.181 |
| 250.037 | 10.02226 | 96.137 | 0.034 | 0.035 | −0.003 | −0.020 | −0.168 |
| 250.037 | 9.01596 | 84.522 | 0.033 | 0.038 | −0.009 | 0.002 | −0.155 |
| 250.038 | 8.01353 | 73.224 | 0.031 | 0.043 | −0.022 | 0.016 | −0.144 |
| 250.038 | 7.00732 | 62.285 | 0.030 | 0.048 | −0.037 | 0.024 | −0.133 |
| 250.039 | 6.00617 | 51.876 | 0.029 | 0.055 | −0.056 | 0.020 | −0.127 |
| 250.041 | 5.00456 | 41.980 | 0.028 | 0.066 | −0.074 | 0.007 | −0.123 |
| 250.038 | 4.00336 | 32.615 | 0.027 | 0.082 | −0.090 | −0.013 | −0.122 |
| 250.038 | 2.98984 | 23.663 | 0.026 | 0.108 | −0.071 | −0.005 | −0.089 |
| 250.037 | 2.00040 | 15.401 | 0.025 | 0.160 | −0.041 | 0.007 | −0.050 |
| 250.041 | 1.00007 | 7.500 | 0.024 | 0.316 | 0.088 | 0.115 | 0.085 |
| | | | | 275.000 K | | | |
| 275.040 | 19.52291 | 158.425 | 0.041 | 0.026 | −0.092 | −0.077 | −0.058 |
| 275.041 | 19.06379 | 155.138 | 0.041 | 0.026 | −0.087 | −0.076 | −0.065 |
| 275.042 | 18.04784 | 147.614 | 0.040 | 0.027 | −0.075 | −0.074 | −0.079 |
| 275.041 | 17.03708 | 139.792 | 0.039 | 0.028 | −0.065 | −0.072 | −0.088 |
| 275.041 | 16.03514 | 131.721 | 0.038 | 0.029 | −0.057 | −0.069 | −0.094 |
| 275.041 | 15.03326 | 123.370 | 0.037 | 0.030 | −0.048 | −0.064 | −0.095 |
| 275.042 | 14.02489 | 114.721 | 0.036 | 0.031 | −0.041 | −0.057 | −0.094 |
| 275.042 | 13.02105 | 105.923 | 0.035 | 0.033 | −0.035 | −0.050 | −0.093 |
| 275.041 | 12.01916 | 97.016 | 0.034 | 0.035 | −0.033 | −0.044 | −0.093 |
| 275.040 | 11.01746 | 88.058 | 0.033 | 0.037 | −0.027 | −0.031 | −0.087 |
| 275.040 | 10.01451 | 79.091 | 0.032 | 0.040 | −0.031 | −0.027 | −0.090 |
| 275.040 | 9.01211 | 70.208 | 0.031 | 0.044 | −0.035 | −0.021 | −0.090 |
| 275.040 | 8.00822 | 61.445 | 0.030 | 0.049 | −0.041 | −0.017 | −0.091 |
| 275.042 | 7.00789 | 52.898 | 0.029 | 0.055 | −0.042 | −0.010 | −0.086 |
| 275.042 | 6.00482 | 44.545 | 0.028 | 0.063 | −0.046 | −0.009 | −0.082 |
| 275.041 | 5.00325 | 36.450 | 0.027 | 0.074 | −0.048 | −0.009 | −0.077 |
| 275.041 | 4.00231 | 28.623 | 0.026 | 0.091 | −0.045 | −0.007 | −0.066 |
| 275.040 | 2.98906 | 20.977 | 0.025 | 0.120 | −0.026 | 0.007 | −0.040 |
| 275.041 | 2.00066 | 13.782 | 0.024 | 0.177 | −0.007 | 0.018 | −0.016 |
| 275.040 | 1.00071 | 6.767 | 0.024 | 0.349 | 0.039 | 0.054 | 0.035 |
| | | | | 300.000 K | | | |
| 300.080 | 19.86739 | 137.734 | 0.039 | 0.028 | −0.021 | −0.012 | −0.017 |
| 300.079 | 19.01352 | 132.264 | 0.038 | 0.029 | −0.019 | −0.013 | −0.017 |
| 300.080 | 18.01111 | 125.661 | 0.037 | 0.030 | −0.017 | −0.012 | −0.016 |
| 300.078 | 17.01112 | 118.891 | 0.036 | 0.031 | −0.016 | −0.013 | −0.015 |
| 300.078 | 16.01481 | 111.982 | 0.036 | 0.032 | −0.014 | −0.013 | −0.014 |
| 300.078 | 15.01186 | 104.882 | 0.035 | 0.033 | −0.014 | −0.013 | −0.014 |
| 300.078 | 14.01252 | 97.691 | 0.034 | 0.035 | −0.011 | −0.011 | −0.014 |
| 300.079 | 13.00993 | 90.386 | 0.033 | 0.037 | −0.010 | −0.009 | −0.015 |
| 300.078 | 12.00725 | 83.021 | 0.032 | 0.039 | −0.009 | −0.007 | −0.017 |
| 300.078 | 11.00800 | 75.658 | 0.031 | 0.042 | −0.003 | < 0.001 | −0.016 |
| 300.078 | 10.00623 | 68.274 | 0.031 | 0.045 | −0.006 | 0.001 | −0.021 |
| 300.079 | 9.00729 | 60.948 | 0.030 | 0.049 | −0.006 | 0.003 | −0.024 |
| 300.078 | 8.00532 | 53.664 | 0.029 | 0.054 | −0.006 | 0.006 | −0.025 |
| 300.078 | 7.00595 | 46.488 | 0.028 | 0.061 | −0.004 | 0.011 | −0.023 |
| 300.077 | 6.00506 | 39.409 | 0.027 | 0.069 | −0.005 | 0.012 | −0.022 |
| 300.076 | 5.00463 | 32.462 | 0.027 | 0.082 | −0.005 | 0.013 | −0.020 |
| 300.077 | 4.00398 | 25.655 | 0.026 | 0.101 | −0.004 | 0.014 | −0.017 |
| 300.078 | 3.00382 | 19.004 | 0.025 | 0.132 | −0.001 | 0.015 | −0.011 |
| 300.076 | 2.00305 | 12.509 | 0.024 | 0.194 | 0.005 | 0.018 | −0.001 |
| 300.076 | 1.00193 | 6.175 | 0.024 | 0.382 | 0.018 | 0.027 | 0.016 |
| | | | | 325.000 K | | | |
| 325.081 | 19.89360 | 121.127 | 0.037 | 0.030 | −0.023 | −0.012 | −0.004 |
| 325.082 | 19.01219 | 116.089 | 0.036 | 0.031 | −0.024 | −0.013 | −0.001 |
| 325.081 | 18.01151 | 110.255 | 0.035 | 0.032 | −0.025 | −0.015 | < 0.001 |
| 325.084 | 17.01191 | 104.312 | 0.035 | 0.033 | −0.024 | −0.015 | 0.002 |
| 325.082 | 16.01309 | 98.270 | 0.034 | 0.035 | −0.026 | −0.018 | < 0.001 |
| 325.081 | 15.01025 | 92.117 | 0.033 | 0.036 | −0.025 | −0.018 | < 0.001 |
| 325.081 | 14.00797 | 85.892 | 0.033 | 0.038 | −0.024 | −0.018 | −0.002 |
| 325.081 | 13.00865 | 79.626 | 0.032 | 0.040 | −0.022 | −0.017 | −0.004 |
| 325.081 | 12.00826 | 73.311 | 0.031 | 0.043 | −0.022 | −0.018 | −0.008 |
| 325.079 | 11.00604 | 66.966 | 0.031 | 0.046 | −0.015 | −0.011 | −0.006 |
| 325.081 | 10.00712 | 60.628 | 0.030 | 0.049 | −0.015 | −0.011 | −0.010 |
| 325.080 | 9.00677 | 54.294 | 0.029 | 0.054 | −0.015 | −0.010 | −0.014 |
| 325.079 | 8.00716 | 47.990 | 0.028 | 0.059 | −0.016 | −0.010 | −0.018 |
| 325.079 | 7.00597 | 41.721 | 0.028 | 0.066 | −0.013 | −0.007 | −0.017 |
| 325.081 | 6.00383 | 35.499 | 0.027 | 0.076 | −0.016 | −0.008 | −0.021 |
| 325.082 | 5.00406 | 29.362 | 0.026 | 0.089 | −0.017 | −0.009 | −0.022 |
| 325.084 | 4.00309 | 23.297 | 0.026 | 0.110 | −0.019 | −0.011 | −0.024 |
| 325.084 | 3.00294 | 17.324 | 0.025 | 0.143 | −0.033 | −0.024 | −0.036 |
| 325.084 | 2.00276 | 11.450 | 0.024 | 0.211 | −0.038 | −0.030 | −0.039 |
| 325.077 | 1.00145 | 5.672 | 0.024 | 0.415 | −0.041 | −0.034 | −0.040 |
| | | | | 350.000 K | | | |
| 350.077 | 19.87136 | 108.394 | 0.035 | 0.032 | −0.022 | −0.009 | 0.012 |
| 350.079 | 19.00817 | 103.959 | 0.035 | 0.033 | −0.023 | −0.010 | 0.013 |
| 350.078 | 18.00707 | 98.733 | 0.034 | 0.035 | −0.026 | −0.013 | 0.013 |
| 350.079 | 17.00650 | 93.431 | 0.034 | 0.036 | −0.026 | −0.014 | 0.013 |
| 350.079 | 16.00701 | 88.062 | 0.033 | 0.037 | −0.025 | −0.015 | 0.014 |
| 350.077 | 15.00673 | 82.624 | 0.032 | 0.039 | −0.025 | −0.016 | 0.012 |
| 350.075 | 14.00468 | 77.121 | 0.032 | 0.041 | −0.024 | −0.017 | 0.009 |
| 350.075 | 13.00594 | 71.590 | 0.031 | 0.043 | −0.022 | −0.017 | 0.007 |
| 350.075 | 12.00520 | 66.012 | 0.030 | 0.046 | −0.020 | −0.016 | 0.005 |
| 350.074 | 11.00582 | 60.421 | 0.030 | 0.049 | −0.012 | −0.009 | 0.008 |
| 350.075 | 10.00544 | 54.804 | 0.029 | 0.053 | −0.013 | −0.011 | 0.003 |
| 350.074 | 9.00462 | 49.183 | 0.028 | 0.058 | −0.013 | −0.011 | −0.001 |
| 350.073 | 8.00495 | 43.575 | 0.028 | 0.064 | −0.014 | −0.012 | −0.005 |
| 350.075 | 7.00384 | 37.977 | 0.027 | 0.072 | −0.009 | −0.006 | −0.002 |
| 350.075 | 6.00276 | 32.405 | 0.027 | 0.082 | −0.011 | −0.007 | −0.005 |
| 350.076 | 5.00261 | 26.873 | 0.026 | 0.096 | −0.011 | −0.007 | −0.007 |
| 350.075 | 4.00245 | 21.385 | 0.025 | 0.118 | −0.015 | −0.009 | −0.011 |
| 350.075 | 3.00247 | 15.949 | 0.025 | 0.155 | −0.022 | −0.016 | −0.019 |
| 350.074 | 2.00193 | 10.569 | 0.024 | 0.228 | −0.028 | −0.022 | −0.025 |
| 350.072 | 1.00211 | 5.257 | 0.023 | 0.446 | −0.014 | −0.008 | −0.011 |

(a) Expanded uncertainties ($k$ = 2): $U(p > 3)/\mathrm{MPa} = 75 \cdot 10^{-6} \cdot \frac{p}{\mathrm{MPa}} + 3.5 \cdot 10^{-3}$; $U(p < 3)/\mathrm{MPa} = 60 \cdot 10^{-6} \cdot \frac{p}{\mathrm{MPa}} + 1.7 \cdot 10^{-3}$; $U(T)$ = 15 mK; $\frac{U(\rho)}{\mathrm{kg \cdot m^{-3}}} = 2.5 \cdot 10^{4} \frac{\chi_s}{\mathrm{m^3 kg^{-1}}} + 1.1 \cdot 10^{-4} \cdot \frac{\rho}{\mathrm{kg \cdot m^{-3}}} + 2.3 \cdot 10^{-2}$.

**Table 7.** Experimental ($p$, $\rho_{exp}$, $T$) measurements for the ($H_2$-enriched) natural gas mixture G 454 (G 431 + 20 % $H_2$), absolute and relative expanded ($k$ = 2) uncertainty in density, $U(\rho_{exp})$, relative deviations from the density given by the AGA8-DC92 EoS [16], $\rho_{AGA8\text{-}DC92}$, the GERG-2008 EoS [17,18], $\rho_{GERG\text{-}2008}$, and the improved GERG-2008 EoS [41–43], $\rho_{GERG\text{-}improved}$.

| $T$ / K[(a)] | $p$ / MPa[(a)] | $\rho_{exp}$ / kg·m$^{-3}$[(a)] | $U(\rho_{exp})$ / kg·m$^{-3}$ | $10^2$ $U(\rho_{exp})/\rho_{exp}$ | $10^2$ ($\rho_{exp}$ − $\rho_{AGA8\text{-}DC92}$)/$\rho_{AGA8\text{-}DC92}$ | $10^2$ ($\rho_{exp}$ − $\rho_{GERG\text{-}2008}$)/ $\rho_{GERG\text{-}2008}$ | $10^2$ ($\rho_{exp}$ − $\rho_{GERG\text{-}improved}$)/$\rho_{GERG\text{-}improved}$ |
|---|---|---|---|---|---|---|---|
| | | | | 250.000 K | | | |
| 250.134 | 19.85938 | 160.136 | 0.041 | 0.026 | −0.119 | −0.108 | −0.123 |
| 250.133 | 19.05712 | 154.693 | 0.040 | 0.026 | −0.089 | −0.099 | −0.146 |
| 250.134 | 18.06430 | 147.637 | 0.040 | 0.027 | −0.053 | −0.085 | −0.167 |
| 250.132 | 17.03403 | 139.940 | 0.039 | 0.028 | −0.019 | −0.070 | −0.181 |
| 250.132 | 16.04267 | 132.179 | 0.038 | 0.029 | 0.008 | −0.054 | −0.186 |
| 250.130 | 15.04385 | 124.038 | 0.037 | 0.030 | 0.035 | −0.032 | −0.179 |
| 250.131 | 14.03791 | 115.535 | 0.036 | 0.031 | 0.056 | −0.011 | −0.167 |
| 250.133 | 13.03272 | 106.796 | 0.035 | 0.033 | 0.077 | 0.018 | −0.144 |
| 250.132 | 12.03336 | 97.915 | 0.034 | 0.035 | 0.080 | 0.035 | −0.131 |
| 250.135 | 11.02191 | 88.824 | 0.033 | 0.037 | 0.085 | 0.060 | −0.111 |
| 250.133 | 10.02295 | 79.806 | 0.032 | 0.040 | 0.071 | 0.070 | −0.106 |
| 250.133 | 9.01852 | 70.792 | 0.031 | 0.044 | 0.056 | 0.080 | −0.100 |
| 250.133 | 8.01228 | 61.881 | 0.030 | 0.048 | 0.030 | 0.078 | −0.102 |
| 250.131 | 7.01013 | 53.193 | 0.029 | 0.054 | 0.007 | 0.074 | −0.101 |
| 250.131 | 6.00760 | 44.730 | 0.028 | 0.063 | −0.021 | 0.059 | −0.104 |
| 250.132 | 5.00600 | 36.542 | 0.027 | 0.074 | −0.046 | 0.040 | −0.106 |
| 250.133 | 4.00423 | 28.641 | 0.026 | 0.091 | −0.060 | 0.022 | −0.101 |
| 250.149 | 3.00373 | 21.046 | 0.025 | 0.120 | −0.063 | 0.008 | −0.087 |
| 250.147 | 2.00285 | 13.751 | 0.024 | 0.178 | −0.029 | 0.025 | −0.040 |
| 250.147 | 1.00235 | 6.747 | 0.024 | 0.350 | 0.066 | 0.096 | 0.063 |
| | | | | 275.000 K | | | |
| 275.111 | 19.79474 | 135.100 | 0.038 | 0.028 | −0.057 | −0.053 | −0.114 |
| 275.105 | 18.04682 | 124.117 | 0.037 | 0.030 | −0.043 | −0.049 | −0.107 |
| 275.105 | 17.04594 | 117.552 | 0.036 | 0.031 | −0.034 | −0.044 | −0.099 |
| 275.104 | 16.03762 | 110.752 | 0.035 | 0.032 | −0.028 | −0.042 | −0.092 |
| 275.104 | 15.03508 | 103.826 | 0.035 | 0.033 | −0.023 | −0.039 | −0.086 |
| 275.102 | 14.03092 | 96.750 | 0.034 | 0.035 | −0.021 | −0.037 | −0.083 |
| 275.085 | 13.02770 | 89.570 | 0.033 | 0.037 | −0.030 | −0.043 | −0.091 |
| 275.094 | 12.02608 | 82.322 | 0.032 | 0.039 | −0.026 | −0.034 | −0.087 |
| 275.098 | 11.01666 | 74.976 | 0.031 | 0.042 | −0.022 | −0.024 | −0.082 |
| 275.100 | 10.01570 | 67.674 | 0.031 | 0.045 | −0.031 | −0.025 | −0.090 |
| 275.097 | 9.01262 | 60.383 | 0.030 | 0.049 | −0.042 | −0.027 | −0.099 |
| 275.096 | 8.01109 | 53.160 | 0.029 | 0.054 | −0.053 | −0.029 | −0.106 |
| 275.096 | 7.00908 | 46.020 | 0.028 | 0.061 | −0.059 | −0.029 | −0.107 |
| 275.098 | 6.00648 | 38.986 | 0.027 | 0.070 | −0.064 | −0.029 | −0.104 |
| 275.098 | 5.00590 | 32.097 | 0.027 | 0.083 | −0.067 | −0.029 | −0.098 |
| 275.089 | 4.00439 | 25.349 | 0.026 | 0.102 | −0.065 | −0.028 | −0.089 |
| 275.098 | 3.00345 | 18.761 | 0.025 | 0.133 | −0.056 | −0.022 | −0.070 |
| 275.097 | 2.00254 | 12.341 | 0.024 | 0.197 | −0.029 | −0.003 | −0.036 |
| 275.094 | 1.00209 | 6.093 | 0.024 | 0.387 | 0.026 | 0.042 | 0.024 |
| | | | | 300.000 K | | | |
| 300.065 | 19.88020 | 118.054 | 0.036 | 0.031 | −0.030 | −0.024 | −0.034 |
| 300.062 | 19.02143 | 113.328 | 0.036 | 0.032 | −0.029 | −0.024 | −0.028 |
| 300.064 | 18.02455 | 107.722 | 0.035 | 0.033 | −0.026 | −0.022 | −0.020 |
| 300.065 | 17.01538 | 101.923 | 0.034 | 0.034 | −0.024 | −0.022 | −0.014 |
| 300.064 | 16.01979 | 96.095 | 0.034 | 0.035 | −0.022 | −0.022 | −0.011 |
| 300.065 | 15.01664 | 90.124 | 0.033 | 0.037 | −0.019 | −0.020 | −0.008 |
| 300.065 | 14.01926 | 84.105 | 0.032 | 0.039 | −0.017 | −0.020 | −0.009 |
| 300.065 | 13.01358 | 77.969 | 0.032 | 0.041 | −0.016 | −0.019 | −0.012 |
| 300.064 | 12.01564 | 71.830 | 0.031 | 0.043 | −0.016 | −0.018 | −0.016 |
| 300.062 | 11.01460 | 65.644 | 0.030 | 0.046 | −0.010 | −0.010 | −0.016 |
| 300.061 | 10.00901 | 59.408 | 0.030 | 0.050 | −0.015 | −0.013 | −0.025 |
| 300.060 | 9.01290 | 53.239 | 0.029 | 0.054 | −0.013 | −0.008 | −0.027 |
| 300.061 | 8.00924 | 47.042 | 0.028 | 0.060 | −0.015 | −0.007 | −0.031 |
| 300.059 | 7.00775 | 40.898 | 0.028 | 0.067 | −0.014 | −0.003 | −0.031 |
| 300.062 | 6.00537 | 34.800 | 0.027 | 0.077 | −0.014 | < 0.001 | −0.029 |
| 300.061 | 5.00462 | 28.778 | 0.026 | 0.091 | −0.014 | 0.002 | −0.026 |
| 300.061 | 4.00319 | 22.830 | 0.025 | 0.112 | −0.013 | 0.003 | −0.021 |
| 300.062 | 3.00277 | 16.975 | 0.025 | 0.146 | −0.012 | 0.004 | −0.017 |
| 300.062 | 2.00253 | 11.219 | 0.024 | 0.215 | 0.007 | 0.020 | 0.006 |
| 300.063 | 1.00253 | 5.564 | 0.023 | 0.422 | 0.033 | 0.042 | 0.034 |
| | | | | 325.000 K | | | |
| 325.068 | 19.91762 | 105.164 | 0.035 | 0.033 | −0.029 | −0.022 | 0.004 |
| 325.070 | 19.03166 | 100.796 | 0.034 | 0.034 | −0.030 | −0.023 | 0.008 |
| 325.072 | 18.02626 | 95.755 | 0.034 | 0.035 | −0.032 | −0.026 | 0.010 |
| 325.075 | 17.02502 | 90.653 | 0.033 | 0.037 | −0.031 | −0.027 | 0.012 |
| 325.074 | 16.02207 | 85.467 | 0.033 | 0.038 | −0.031 | −0.028 | 0.012 |
| 325.078 | 15.01620 | 80.196 | 0.032 | 0.040 | −0.030 | −0.028 | 0.010 |
| 325.080 | 14.01608 | 74.897 | 0.031 | 0.042 | −0.028 | −0.028 | 0.008 |
| 325.081 | 13.01381 | 69.537 | 0.031 | 0.044 | −0.027 | −0.028 | 0.004 |
| 325.081 | 12.01465 | 64.153 | 0.030 | 0.047 | −0.026 | −0.028 | −0.001 |
| 325.080 | 11.01158 | 58.724 | 0.030 | 0.050 | −0.018 | −0.021 | 0.001 |
| 325.080 | 10.00777 | 53.265 | 0.029 | 0.054 | −0.020 | −0.022 | −0.007 |
| 325.083 | 9.01024 | 47.834 | 0.028 | 0.059 | −0.017 | −0.019 | −0.009 |
| 325.080 | 8.00638 | 42.370 | 0.028 | 0.065 | −0.021 | −0.021 | −0.016 |
| 325.076 | 7.00438 | 36.931 | 0.027 | 0.073 | −0.019 | −0.017 | −0.016 |
| 325.076 | 6.00516 | 31.529 | 0.026 | 0.084 | −0.021 | −0.017 | −0.019 |
| 325.078 | 5.00390 | 26.146 | 0.026 | 0.099 | −0.024 | −0.018 | −0.022 |
| 325.077 | 4.00361 | 20.810 | 0.025 | 0.121 | −0.026 | −0.020 | −0.023 |
| 325.081 | 3.00304 | 15.519 | 0.025 | 0.159 | −0.035 | −0.027 | −0.031 |
| 325.076 | 2.00278 | 10.288 | 0.024 | 0.234 | −0.036 | −0.028 | −0.031 |
| 325.078 | 1.00254 | 5.117 | 0.023 | 0.458 | −0.035 | −0.029 | −0.031 |
| | | | | 350.000 K | | | |
| 350.086 | 19.91259 | 95.026 | 0.034 | 0.035 | −0.030 | −0.025 | 0.017 |
| 350.083 | 19.01409 | 91.018 | 0.033 | 0.037 | −0.032 | −0.027 | 0.019 |
| 350.088 | 18.01116 | 86.481 | 0.033 | 0.038 | −0.030 | −0.026 | 0.022 |
| 350.089 | 17.01022 | 81.891 | 0.032 | 0.039 | −0.030 | −0.027 | 0.023 |
| 350.090 | 16.00870 | 77.239 | 0.032 | 0.041 | −0.028 | −0.027 | 0.023 |
| 350.091 | 15.01019 | 72.551 | 0.031 | 0.043 | −0.025 | −0.025 | 0.024 |
| 350.092 | 14.00745 | 67.794 | 0.031 | 0.045 | −0.023 | −0.025 | 0.022 |
| 350.093 | 13.00789 | 63.011 | 0.030 | 0.048 | −0.021 | −0.024 | 0.019 |
| 350.091 | 12.00843 | 58.193 | 0.029 | 0.051 | −0.019 | −0.023 | 0.016 |
| 350.092 | 11.00574 | 53.335 | 0.029 | 0.054 | −0.011 | −0.015 | 0.019 |
| 350.092 | 10.00748 | 48.471 | 0.028 | 0.059 | −0.011 | −0.015 | 0.015 |
| 350.091 | 9.00560 | 43.576 | 0.028 | 0.064 | −0.012 | −0.015 | 0.010 |
| 350.095 | 8.00468 | 38.677 | 0.027 | 0.071 | −0.010 | −0.012 | 0.009 |
| 350.096 | 7.00500 | 33.785 | 0.027 | 0.079 | −0.008 | −0.009 | 0.008 |
| 350.094 | 6.00374 | 28.888 | 0.026 | 0.091 | −0.012 | −0.012 | 0.003 |
| 350.094 | 5.00298 | 24.006 | 0.026 | 0.107 | −0.018 | −0.016 | −0.004 |
| 350.093 | 4.00375 | 19.151 | 0.025 | 0.131 | −0.024 | −0.020 | −0.011 |
| 350.092 | 3.00275 | 14.311 | 0.024 | 0.171 | −0.038 | −0.032 | −0.025 |
| 350.094 | 2.00271 | 9.507 | 0.024 | 0.252 | −0.051 | −0.045 | −0.040 |
| 350.094 | 1.00247 | 4.739 | 0.023 | 0.494 | −0.057 | −0.051 | −0.049 |

[a] Expanded uncertainties ($k$ = 2): $U(p > 3)/\mathrm{MPa} = 75 \cdot 10^{-6} \cdot \frac{p}{\mathrm{MPa}} + 3.5 \cdot 10^{-3}$; $U(p < 3)/\mathrm{MPa} = 60 \cdot 10^{-6} \cdot \frac{p}{\mathrm{MPa}} + 1.7 \cdot 10^{-3}$; $U(T)$ = 15 mK; $\frac{U(\rho)}{\mathrm{kg \cdot m^{-3}}} = 2.5 \cdot 10^{4} \frac{\chi_S}{\mathrm{m^3 kg^{-1}}} + 1.1 \cdot 10^{-4} \cdot \frac{\rho}{\mathrm{kg \cdot m^{-3}}} + 2.3 \cdot 10^{-2}$.

**Table 8.** Statistical analysis of the (*p*, *ρ*, *T*) data sets with respect to AGA8-DC92 EoS [16], GERG-2008 EoS [17,18], and improved GERG-2008 EoS [41–43] for all the natural gas mixtures studied in this work, including literature data for comparable mixtures. AARD = average absolute value of the relative deviations, BiasRD = average relative deviation, RMSRD = root mean square relative deviation, MaxRD = maximum value of the relative deviations.

| Reference[(a)] | $x_{H_2}$ | $N$[(b)] | Covered ranges | | Experimental *vs* AGA8-DC92 EoS | | | | Experimental *vs* GERG-2008 EoS | | | | Experimental *vs* improved GERG-2008 EoS | | | |
|---|---|---|---|---|---|---|---|---|---|---|---|---|---|---|---|---|
| | | | $T$ / K | $p$ / MPa | AARD / % | Bias / % | RMS / % | MaxRD / % | AARD / % | Bias / % | RMS / % | MaxRD / % | AARD / % | Bias / % | RMS / % | MaxRD / % |
| G 431 (this work) | 0 | 97 | 250–350 | 1–20 | 0.012 | −0.0066 | 0.018 | 0.054 | 0.032 | 0.032 | 0.034 | 0.049 | 0.019 | 0.016 | 0.023 | 0.051 |
| G 453 (this work) | 0.099928 | 98 | 275–350 | 1–20 | 0.032 | −0.029 | 0.045 | 0.19 | 0.029 | −0.021 | 0.049 | 0.18 | 0.047 | −0.042 | 0.070 | 0.20 |
| G 454 (this work) | 0.199945 | 99 | 250–350 | 1–20 | 0.033 | −0.020 | 0.039 | 0.12 | 0.030 | −0.015 | 0.037 | 0.11 | 0.052 | −0.043 | 0.072 | 0.19 |
| Hernández-Gómez et al., 2018 [24] | 0.030097 | 99 | 260–350 | 1–20 | 0.066 | −0.066 | 0.074 | 0.20 | 0.10 | −0.10 | 0.13 | 0.31 | 0.13 | −0.12 | 0.15 | 0.33 |
| Richter et al., 2014 [47][(c)] | 0.053681 | 37 | 273–293 | 1–8 | 0.059 | −0.059 | 0.061 | 0.078 | 0.041 | −0.041 | 0.044 | 0.075 | 0.094 | −0.094 | 0.10 | 0.14 |
| Richter et al., 2014 [47][(c)] | 0.104106 | 36 | 273–293 | 1–8 | 0.019 | −0.010 | 0.021 | 0.039 | 0.021 | 0.019 | 0.026 | 0.055 | 0.054 | −0.054 | 0.061 | 0.10 |
| Richter et al., 2014 [47][(c)] | 0.304705 | 13 | 283 | 1–8 | 0.26 | −0.26 | 0.26 | 0.27 | 0.21 | −0.21 | 0.21 | 0.27 | 0.26 | −0.26 | 0.27 | 0.32 |

[(a)] Only vapor and supercritical phase measurements have been considered.

[(b)] Number of experimental points.

[(c)] Only experimental mass densities were considered.

**Table 9.** Derived isothermal compressibility $\kappa_T$ values for all the reference natural gas mixtures studied in this work at various temperatures $T$ and pressures $p$.

| $\kappa_T$ / MPa$^{-1}$ [(a)] | | | | | |
|---|---|---|---|---|---|
| | $T$ / K | | | | |
| $p$ / MPa | 250 | 275 | 300 | 325 | 350 |
| | | | G 431 | | |
| 19 | 0.0335 | | 0.0480 | 0.0495 | 0.0497 |
| 18 | 0.0378 | 0.0486 | 0.0530 | 0.0540 | 0.0539 |
| 17 | 0.0435 | 0.0548 | 0.0581 | 0.0583 | 0.0579 |
| 16 | 0.0509 | 0.0615 | 0.0636 | 0.0632 | 0.0625 |
| 15 | 0.0599 | 0.0692 | 0.0698 | 0.0686 | 0.0675 |
| 14 | 0.0711 | 0.0776 | 0.0766 | 0.0746 | 0.0730 |
| 13 | 0.0844 | 0.0869 | 0.0840 | 0.0813 | 0.0793 |
| 12 | 0.0995 | 0.0970 | 0.0923 | 0.0889 | 0.0865 |
| 11 | 0.1160 | 0.1079 | 0.1016 | 0.0976 | 0.0949 |
| 10 | 0.1329 | 0.1199 | 0.1123 | 0.1077 | 0.1048 |
| 9 | 0.1497 | 0.1333 | 0.1247 | 0.1197 | 0.1165 |
| 8 | 0.1670 | 0.1489 | 0.1397 | 0.1345 | 0.1311 |
| 7 | 0.1861 | 0.1678 | 0.1585 | 0.1531 | 0.1495 |
| 6 | 0.2097 | 0.1923 | 0.1830 | 0.1775 | 0.1739 |
| 5 | 0.2418 | 0.2258 | 0.2169 | 0.2114 | 0.2077 |
| 4 | 0.2900 | 0.2757 | 0.2672 | 0.2618 | 0.2580 |
| 3 | 0.3711 | 0.3590 | 0.3509 | 0.3457 | 0.3423 |
| 2 | 0.5361 | 0.5215 | 0.5142 | 0.5088 | 0.5001 |
| | | | G 453 (G 431 + 10 % $H_2$) | | |
| 19 | | 0.0468 | 0.0488 | 0.0494 | 0.0493 |
| 18 | | 0.0515 | 0.0533 | 0.0535 | 0.0534 |
| 17 | 0.0519 | 0.0567 | 0.0577 | 0.0575 | 0.0571 |
| 16 | 0.0585 | 0.0624 | 0.0627 | 0.0621 | 0.0614 |
| 15 | 0.0664 | 0.0688 | 0.0682 | 0.0671 | 0.0662 |
| 14 | 0.0754 | 0.0758 | 0.0743 | 0.0727 | 0.0716 |
| 13 | 0.0854 | 0.0836 | 0.0811 | 0.0791 | 0.0777 |
| 12 | 0.0963 | 0.0922 | 0.0887 | 0.0863 | 0.0846 |
| 11 | 0.1082 | 0.1019 | 0.0976 | 0.0947 | 0.0928 |
| 10 | 0.1212 | 0.1129 | 0.1078 | 0.1047 | 0.1026 |
| 9 | 0.1354 | 0.1256 | 0.1200 | 0.1165 | 0.1142 |
| 8 | 0.1516 | 0.1408 | 0.1348 | 0.1311 | 0.1286 |
| 7 | 0.1710 | 0.1598 | 0.1535 | 0.1496 | 0.1470 |
| 6 | 0.1956 | 0.1845 | 0.1781 | 0.1741 | 0.1713 |
| 5 | 0.2292 | 0.2184 | 0.2120 | 0.2079 | 0.2051 |
| 4 | 0.2789 | 0.2686 | 0.2624 | 0.2585 | 0.2554 |
| 3 | 0.3614 | 0.3522 | 0.3463 | 0.3424 | 0.3397 |
| 2 | 0.5286 | 0.5165 | 0.5106 | 0.5064 | 0.4989 |
| | | | G 454 (G 431 + 20 % $H_2$) | | |
| 19 | 0.0452 | 0.0479 | 0.0489 | 0.0491 | 0.0489 |
| 18 | 0.0497 | 0.0520 | 0.0529 | 0.0529 | 0.0527 |
| 17 | 0.0549 | 0.0568 | 0.0570 | 0.0567 | 0.0564 |
| 16 | 0.0608 | 0.0618 | 0.0616 | 0.0610 | 0.0605 |
| 15 | 0.0673 | 0.0674 | 0.0666 | 0.0658 | 0.0651 |
| 14 | 0.0746 | 0.0737 | 0.0723 | 0.0712 | 0.0703 |
| 13 | 0.0827 | 0.0806 | 0.0787 | 0.0773 | 0.0763 |
| 12 | 0.0917 | 0.0884 | 0.0860 | 0.0843 | 0.0831 |
| 11 | 0.1018 | 0.0974 | 0.0945 | 0.0925 | 0.0912 |
| 10 | 0.1132 | 0.1079 | 0.1045 | 0.1023 | 0.1008 |
| 9 | 0.1263 | 0.1202 | 0.1164 | 0.1140 | 0.1123 |
| 8 | 0.1419 | 0.1352 | 0.1310 | 0.1285 | 0.1266 |
| 7 | 0.1611 | 0.1540 | 0.1496 | 0.1468 | 0.1449 |
| 6 | 0.1860 | 0.1786 | 0.1741 | 0.1712 | 0.1692 |
| 5 | 0.2200 | 0.2126 | 0.2080 | 0.2050 | 0.2029 |
| 4 | 0.2703 | 0.2631 | 0.2585 | 0.2555 | 0.2532 |
| 3 | 0.3533 | 0.3470 | 0.3424 | 0.3394 | 0.3374 |
| 2 | 0.5215 | 0.5121 | 0.5073 | 0.5040 | 0.4975 |

[(a)] Expanded ($k$ = 2) uncertainty: $U_r(\kappa_T)$ = 0.7 %.